\documentclass[10pt,twocolumn,twoside,a4paper]{IEEEtran}

\usepackage{cite}
\usepackage{amsmath,amssymb,amsfonts}
\usepackage{algorithmic}
\usepackage{graphicx}
\usepackage{textcomp}
\usepackage{booktabs} % For formal tables
\usepackage[utf8]{inputenc}
\usepackage[T1]{fontenc}
\usepackage{float}
\usepackage{amssymb}
\usepackage{amsmath}
\usepackage{subfigure}
\usepackage{color}
\usepackage{multirow}
\usepackage{relsize}

\graphicspath{{figures/}}

\definecolor{amethyst}{rgb}{0.6, 0.4, 0.8}
\definecolor{blue-violet}{rgb}{0.54, 0.17, 0.89}
\definecolor{blue(pigment)}{rgb}{0.2, 0.2, 0.6}
\definecolor{byzantium}{rgb}{0.44, 0.16, 0.39}
\definecolor{coolblack}{rgb}{0.0, 0.18, 0.39}
\definecolor{english}{rgb}{0.0, 0.5, 0.0}
\definecolor{armygreen}{rgb}{0.29, 0.33, 0.13}
\definecolor{red}{rgb}{1,0,0} % Europe - red
\definecolor{yellow}{rgb}{0.8, 0.8, 0} % Middle east - yellow
\definecolor{blue}{rgb}{.05,.05,.9} % Russia - blue
\definecolor{purple}{rgb}{0.5,0,0.5} % China - purple 
\definecolor{green}{rgb}{0,0.5,0} % Bresil south america - green
\definecolor{orange}{rgb}{.8,0.5,0} %EN-US color - orange
\definecolor{pink}{rgb}{.8,0,0.8} % South east asia - pink 

\newcommand\GR{{G_{\rm R}}}
\newcommand\Grr{{G_{rr}}}

\newcommand\Gqr{{G_{\rm qr}}}
\newcommand\Gpr{{G_{\rm pr}}}
\newcommand\Gqrd{{G_{\rm qrd}}}
\newcommand\Gqrnd{{G_{\rm qrnd}}}

\def\BibTeX{{\rm B\kern-.05em{\sc i\kern-.025em b}\kern-.08em
    T\kern-.1667em\lower.7ex\hbox{E}\kern-.125emX}}

\begin{document}

\title{Interactions and influence of world painters
from the reduced Google matrix of Wikipedia networks}

%%%%%%%%%%%%%%%%%%%%%%%%%%%%%%%%%%%%%%%%%%%%%%%%%%%%%%%%%%%%%%%%%%%%%%%%%%%%%%%%%%%%%%}
%\history{Date of publication xxxx 00, 0000, date of current version xxxx 00, 0000.}
%\doi{10.1109/ACCESS.2017.DOI}

%\title{Reduced Google matrix analysis and Art: Painters}
%\title{Interactions and influence of world painters
%from reduced Google matrix of Wikipedia networks}
%\author{\uppercase{Samer El Zant}\authorrefmark{1},
%\uppercase{Katia Jaffr\`es-Runser}\authorrefmark{1},
%\uppercase{Klaus M. Frahm\authorrefmark{2},
%and \\ 
%Dima L. Shepelyansky}.\authorrefmark{2},
%}
%\address[1]{Institut de Recherche en Informatique de Toulouse, Universit\'e de Toulouse, INPT, Toulouse, France, Email: \{samer.elzant, kjr\}@enseeiht.fr}
%\address[2]{Laboratoire de Physique Th\'eorique du CNRS, IRSAMC, Universit\'e de Toulouse, UPS, Toulouse, France, Email: \{dima, klaus\}@irsamc.ups-tlse.fr}
%%%%%%%%%%%%%%%%%%%%%%%%%%%%%%%%%%%%%%%%%%%%%%%%%%%%%%%%%%%%%%%%%%%%

\author{Samer~El~Zant$^1$, Katia~Jaffr\`es-Runser$^1$, Klaus~M.~Frahm$^2$ and Dima~L.~Shepelyansky$^2$
\thanks{(1) Samer El Zant and Katia Jaffr\`es-Runser are with the Institut de Recherche en Informatique de Toulouse, Universit\'e de Toulouse, INPT, Toulouse, France.}
\thanks{(2) Klaus M. Frahm and Dima L. Shepelyansky are with the Laboratoire de Physique Th\'eorique, IRSAMC, Universit\'e de Toulouse, CNRS, UPS, 31062 Toulouse, France.}
}

%%%%%%%%%%%%%%%%%%%%%%%%%%%%%%%%%%%%%%%%%%%%%%%%%%%%%%%%%%%%%%%%
%\author{\IEEEauthorblockN{Samer El Zant, Katia Jaffr\`es-Runser}
%\IEEEauthorblockA{Institut de Recherche\\en Informatique de Toulouse\\ Universit\'e de Toulouse, INPT\\
%Toulouse, France\\
%Email: \{samer.elzant, kjr\}@enseeiht.fr}
%\and
%\IEEEauthorblockN{Klaus M. Frahm, Dima L. Shepelyansky}
%\IEEEauthorblockA{Laboratoire de Physique Th\'eorique\\du CNRS, IRSAMC\\ Universit\'e de Toulouse, INPT\\
%Toulouse, France\\
%Email: \{dima, klaus\}@irsamc.ups-tlse.fr}
%}
%%%%%%%%%%%%%%%%%%%%%%%%%%%%%%%%%%%%%%%%%%%%%%%%%%%%

%%%%%%%%%%%%%%%%%%%%%%%%%%%%%%%%%%%%%%%%%%%%%
%\tfootnote{This paragraph of the first footnote will contain support 
%information, including sponsor and financial support acknowledgment. For 
%example, ``This work was supported in part by the U.S. Department of 
%Commerce under Grant BS123456.''}

%\markboth
%{El Zant \headeretal: Interactions and influence of world painters}
%{El Zant \headeretal: Interactions and influence of world painters}
%{El Zant \headeretal: Reduced Google matrix analysis and Art: Painters}
%{El Zant \headeretal: Reduced Google matrix analysis and Art: Painters}

%\corresp{Corresponding author: Katia Jaffr\`es-Runser (e-mail: kjr@enseeiht.fr).}
%%%%%%%%%%%%%%%%%%%%%%%%%%%%%%%%%%%%%%%%%%%%%%%%%%%%%%%%%%%%%%%%%%%%%%%%%%%%%
\maketitle

\begin{abstract}
This study concentrates on extracting painting art history knowledge 
from the network structure of Wikipedia. Therefore, we construct theoretical networks 
of webpages representing the hyper-linked structure of articles of 7 Wikipedia language editions. 
These 7 networks are analyzed to extract the most influential painters 
in each edition using Google matrix theory. Importance of webpages of over 3000 painters 
are measured using PageRank algorithm. 
The most influential painters are enlisted and their ties are studied with 
the reduced Google matrix analysis. Reduced Google Matrix is a powerful method 
that captures both direct and hidden interactions between a subset of selected nodes
taking into account the indirect links between these nodes via the remaining part of 
large global network. This method originates from the scattering theory of nuclear
and mesoscopic physics and field of quantum chaos. 
From this study, we show that it is possible to extract from the components 
of the reduced Google matrix meaningful information on the ties between these painters. 
For instance, our analysis groups together painters that belong to 
the same painting movement and shows meaningful ties between painters of different movements. 
We also determine the influence of painters on world countries 
using link sensitivity between Wikipedia articles of painters and countries.
The reduced Google matrix approach allows to obtain a balanced
view of various cultural opinions of Wikipedia language editions. 
The world countries with the largest number of 
top painters of selected 7 Wikipedia editions are found to be
Italy, France, Russia.
We argue that this approach gives meaningful information about art and 
that it could be a part of extensive network analysis on human knowledge and cultures.
\end{abstract}


%\titlepgskip=-15pt

%\maketitle

\section{Introduction}
\label{sec:intro}

{\it "The art is the expression or application of human creative skill and imagination, 
typically in a visual form such as painting or sculpture, producing works to be appreciated primarily 
for their beauty or emotional power"}~\cite{oxford}. Artists use different approaches and techniques 
to create emotions. Since the beginning of mankind, painters have offered masterpieces in the form 
of paintings and drawings to the world. 
Depending on historical periods, cultural context and available techniques, painters 
have followed different art movements. Art historians group painters into art movements to capture 
the fact that they have worked in the same school of thought. But a painter could be placed 
in several movements as his works evolve with time and its individual intellectual path 
development~\cite{melivin,fandel,richardson,chipp,richardson2,wolf,benson}.

The major finding of this paper is to show that it is possible to automatically extract this 
common knowledge on art history by analyzing the hyper-linked network structure of the global and 
free online encyclopedia Wikipedia \cite{wikiorg}. The analysis conducted in this work is solely 
based on a graph representation of the Wikipedia articles where vertices (nodes) represent the articles 
and the edges (links) provide the hyperlinks linking these articles together. 
The actual content of articles is never processed in our developments.

%%%%%%%%%%%%%%%%%%%%%%%%%%%%%%%%%%%%%%%%%%%%%%%%%%%%%
\begin{table*}[!ht]
\centering
\caption{List of 50 top painters from FrWiki, EnWiki, DeWiki, ItWiki, EsWiki, NlWiki and RuWiki by increasing PageRank index}
\label{tab:tableall}
\resizebox{\textwidth}{!}{
\begin{tabular}{|c|c|c|c|c|c|c|}
\hline
\textbf{FrWiki}               	&	 \textbf{EnWiki}               	&	 \textbf{DeWiki}          	&	 \textbf{ItWiki}           	&	 \textbf{EsWiki}           	&	 \textbf{NlWiki}           	&	 \textbf{RuWiki}               \\ \hline
Pablo Picasso                 	&	 Leonardo da Vinci             	&	 Leonardo da Vinci        	&	 Leonardo da Vinci         	&	 Leonardo da Vinci         	&	Rembrandt Van Rijn	&	 Leonardo da Vinci             \\ \hline
Leonardo da Vinci             	&	 Pablo Picasso                 	&	 Pablo Picasso            	&	 Michelangelo              	&	 Francisco Goya            	&	Leonardo da Vinci	&	 Pablo Picasso                 \\ \hline
Michelangelo                  	&	 Michelangelo                  	&	 Albrecht Durer           	&	 Raphael                   	&	 Pablo Picasso             	&	Peter Paul Rubens	&	 Michelangelo                  \\ \hline
Claude Monet                  	&	 Raphael                       	&	 Michelangelo             	&	 Pablo Picasso             	&	 Michelangelo              	&	Vincent Van Gogh	&	 Rembrandt Van Rijn            \\ \hline
Vincent Van Gogh              	&	 Rembrandt Van Rijn            	&	 Raphael                  	&	 Giorgio Vasari            	&	 Raphael                   	&	Pablo Picasso	&	 Vincent Van Gogh              \\ \hline
Jacques-Louis David           	&	 Vincent Van Gogh              	&	 Rembrandt Van Rijn       	&	 Titian                    	&	 Diego Vel\'{a}zquez           	&	Johannes Vermeer	&	 Raphael                       \\ \hline
Eug\`{e}ne Delacroix              	&	 Francis Bacon                 	&	 Peter Paul Rubens        	&	 Peter Paul Rubens         	&	 Salvador Dali             	&	Piet Mondrian	&	 Albrecht Durer                \\ \hline
Raphael                       	&	 Andy Warhol                   	&	 Vincent Van Gogh         	&	 Caravaggio                	&	 Peter Paul Rubens         	&	Pieter Bruegel The Elder	&	 Ilya Repin                    \\ \hline
Henri Matisse                 	&	 Peter Paul Rubens             	&	 Titian                   	&	 Vincent Van Gogh          	&	 Titian                    	&	Claude Monet 	&	 Peter Paul Rubens             \\ \hline
Salvador Dali                 	&	 Albrecht Durer                	&	 Francis Bacon            	&	 Giotto Di Bondone         	&	 Francis Bacon             	&	Titian	&	 Nicholas Roerich              \\ \hline
Paul C\'{e}zanne                  	&	 William Blake                 	&	 Andy Warhol              	&	 Rembrandt Van Rijn        	&	 Albrecht Durer            	&	Sandro Botticelli	&	 Titian                        \\ \hline
Rembrandt Van Rijn            	&	 Titian                        	&	 Paul Klee                	&	 Sandro Botticelli         	&	 El Greco                  	&	Paul C\'{e}zanne 	&	 Henri Matisse                 \\ \hline
Peter Paul Rubens             	&	 Claude Monet                  	&	 Paul C\'{e}zanne             	&	 Albrecht Durer            	&	 Rembrandt Van Rijn        	&	Albrecht Durer	&	 Salvador Dali                 \\ \hline
Andy Warhol                   	&	 Salvador Dali                 	&	 Lucas Cranach the Elder  	&	 Francisco Goya            	&	 Vincent Van Gogh          	&	Frans Hals	&	 Paul C\'{e}zanne                  \\ \hline
Marcel Duchamp                	&	 Henri Matisse                 	&	 Wassily Kandinsky        	&	 Giuseppe Arcimboldo       	&	 Sandro Botticelli         	&	Giotto Di Bondone	&	 Viktor Vasnetsov              \\ \hline
\'{E}douard Manet                 	&	 Giorgio Vasari                	&	 Claude Monet             	&	 Piero Della Francesca     	&	 Caravaggio                	&	Jan Van Eyck	&	 Ivan Aivazovsky               \\ \hline
Giorgio Vasari                	&	 Paul C\'{e}zanne                  	&	 Henri Matisse            	&	 Edvard Munch              	&	 Henri Matisse             	&	Andy Warhol	&	 Diego Vel\'{a}zquez               \\ \hline
Paul Gauguin                  	&	 Francisco Goya                	&	 Salvador Dali            	&	 Andrea Mantegna           	&	 Eug\`{e}ne Delacroix          	&	Anthony van Dyck	&	 Marc Chagall                  \\ \hline
Albrecht Durer                	&	 Joseph Mallord William Turner 	&	 Giorgio Vasari           	&	 Masaccio                  	&	 Paul C\'{e}zanne              	&	Paolo Veronese 	&	 Claude Monet                  \\ \hline
Pierre Auguste Renoir         	&	 Eug\`{e}ne Delacroix              	&	 Edvard Munch             	&	 Claude Monet              	&	 Andy Warhol               	&	Francisco Goya	&	 Valentin Serov                \\ \hline
Joan Mir\'{o}                     	&	 Caravaggio                    	&	 Giotto Di Bondone        	&	 Jacques-Louis David       	&	 Claude Monet              	&	Salvador Dali	&	 Paul Gauguin                  \\ \hline
Jean-Auguste-Dominique Ingres 	&	 Jackson Pollock               	&	 Marc Chagall             	&	 Samuel Morse              	&	 Giorgio Vasari            	&	\'{E}douard Manet 	&	 Hieronymus Bosch              \\ \hline
Georges Braque                	&	 \'{E}douard Manet                 	&	 Caspar David Friedrich   	&	 Wassily Kandinsky         	&	 Paul Gauguin              	&	JAMES ENSOR	&	 Henri de Toulouse-Lautrec     \\ \hline
Edgar Degas                   	&	 Anthony van Dyck              	&	 \'{E}douard Manet            	&	 Diego Vel\'{a}zquez           	&	 Diego Rivera              	&	Wassily Kandinsky	&	 Karl Bryullov                 \\ \hline
Francisco Goya                	&	 Pierre Auguste Renoir         	&	 Otto Dix                 	&	 Pieter Bruegel The Elder  	&	 Giotto Di Bondone         	&	Paul Gauguin 	&	 Eug\`{e}ne Delacroix              \\ \hline
Gustave Courbet               	&	 Jacques-Louis David           	&	 Caravaggio               	&	 Fra Angelico              	&	 Jacques-Louis David       	&	Henri Matisse 	&	 Wassily Kandinsky             \\ \hline
Fernand L\'{e}ger                 	&	 Diego Vel\'{a}zquez               	&	 Francisco Goya           	&	 Salvador Dali             	&	 \'{E}douard Manet             	&	William Blake	&	 \'{E}douard Manet                 \\ \hline
Titian                        	&	 William Hogarth               	&	 Pierre Auguste Renoir    	&	 Pierre Auguste Renoir     	&	 Tintoretto                	&	Rene Magritte	&	 Francisco Goya                \\ \hline
Caravaggio                    	&	 Paul Gauguin                  	&	 Paul Gauguin             	&	 Andy Warhol               	&	 Bartolom\'{e} Esteban Murillo 	&	Jacob Jordaens 	&	 Kazimir Malevich              \\ \hline
Jackson Pollock               	&	 Hans Holbein The Younger      	&	 Max Ernst                	&	 Anthony van Dyck          	&	 Anthony van Dyck          	&	Gustav Klimt	&	 Andrei Rublev                 \\ \hline
Wassily Kandinsky             	&	 Edgar Degas                   	&	 Gustav Klimt             	&	 Giovanni Battista Tiepolo 	&	 Georges Braque            	&	Eugène Delacroix 	&	 Giorgio Vasari                \\ \hline
Nicolas Poussin               	&	 Johannes Vermeer              	&	 Eug\`{e}ne Delacroix         	&	 Paul C\'{e}zanne              	&	 Edgar Degas               	&	Karel Appel	&	 Jacques-Louis David           \\ \hline
Marc Chagall                  	&	 Marcel Duchamp                	&	 Joan Mir\'{o}                	&	 Giovanni Bellini          	&	 Joan Mir\'{o}                 	&	Jacques-Louis David 	&	 Igor Grabar                   \\ \hline
Honor\'{e} Daumier                	&	 Sandro Botticelli             	&	 Jan Van Eyck             	&	 Domenico Ghirlandaio      	&	 Wassily Kandinsky         	&	Giorgio Vasari	&	 Pierre Auguste Renoir         \\ \hline
Max Ernst                     	&	 Giotto Di Bondone             	&	 Pieter Bruegel The Elder 	&	 Pietro Perugino           	&	 Hieronymus Bosch          	&	Henry van de Velde 	&	 Samuel Morse                  \\ \hline
Diego Vel\'{a}zquez               	&	 Williem De Kooning            	&	 Max Liebermann           	&	 Jan Van Eyck              	&	 Piero Della Francesca     	&	Henri de Toulouse-Lautrec 	&	 Caravaggio                    \\ \hline
Gustave Dor\'{e}                  	&	 Nicolas Poussin               	&	 Diego Vel\'{a}zquez          	&	 Paolo Veronese            	&	 Andrea Mantegna           	&	Paul Klee	&	 Edgar Degas                   \\ \hline
Sandro Botticelli             	&	 Pieter Bruegel The Elder      	&	 Sandro Botticelli        	&	 Giorgione                 	&	 Jackson Pollock           	&	Marc Chagall 	&	 Mikhail Vrubel                \\ \hline
Giotto Di Bondone             	&	 John Constable                	&	 Marcel Duchamp           	&	 Nicolas Poussin           	&	 Henri de Toulouse-Lautrec 	&	Joseph Mallord William Turner	&	 Nicolas Poussin               \\ \hline
Jean-Baptiste Camille Corot   	&	 Wassily Kandinsky             	&	 Gerhard Richter          	&	 Tintoretto                	&	 Johannes Vermeer          	&	Edvard Munch	&	 Anthony van Dyck              \\ \hline
Henri de Toulouse-Lautrec     	&	 Marc Chagall                  	&	 Max Beckmann             	&	 Paul Gauguin              	&	 Francisco De Zurbaran     	&	Roger Van Der Weyden	&	 Joseph Mallord William Turner \\ \hline
William Bouguereau            	&	 El Greco                      	&	 Hans Holbein The Younger 	&	 Antonio da Correggio      	&	 William Blake             	&	Georges Seurat 	&	 Jean-Auguste-Dominique Ingres \\ \hline
Pieter Bruegel The Elder      	&	 Lucas Cranach the Elder       	&	 El Greco                 	&	 Edgar Degas               	&	 Marcel Duchamp            	&	Nicolas Poussin 	&	 Alexandre Benois              \\ \hline
Antoine Watteau               	&	 Benjamin West                 	&	 Jacques-Louis David      	&	 \'{E}douard Manet             	&	 Pierre Auguste Renoir     	&	Joan Mir\'{o}	&	 Giotto Di Bondone             \\ \hline
Georges Seurat                	&	 Gustave Dor\'{e}                  	&	 Georges Braque           	&	 Lucas Cranach the Elder   	&	 Hans Holbein The Younger  	&	Gustave Dor\'{e} 	&	 Konstantin Korovin            \\ \hline
Rene Magritte                 	&	 Henri de Toulouse-Lautrec     	&	 Johannes Vermeer         	&	 Eug\`{e}ne Delacroix          	&	 Pieter Bruegel The Elder  	&	Edgar Degas 	&	 Isaac Levitan                 \\ \hline
Andr\'{e} Derain                  	&	 Georgia O'keefe               	&	 Henry van de Velde       	&	 Gustave Dor\'{e}              	&	 Nicolas Poussin           	&	Georges Braque 	&	 Gustave Courbet               \\ \hline
Paul Klee                     	&	 James Abbot Mac Neil Whistler 	&	 Edgar Degas              	&	 Marc Chagall              	&	 Jan Van Eyck              	&	Hans Holbein The Younger	&	 William Blake                 \\ \hline
Fran\c{c}ois Boucher              	&	 Jan Van Eyck                  	&	 Lovis Corinth            	&	 Guido Reni                	&	 William Bouguereau        	&	Marcel Duchamp	&	 Tove Jansson                  \\ \hline
Camille Pissarro              	&	 Thomas Gainsborough           	&	 Franz Marc               	&	 William Blake             	&	 Gustave Courbet           	&		&	 Ivan Kramskoi                 \\ \hline
\end{tabular}
}
\end{table*}
%%%%%%%%%%%%%%%%%%%%%%%%%%%%%%%%%%%%%%%%%%%%%%%%%%%%%
Wikipedia has become the largest open source of knowledge being close to Encyclop\ae{}dia Britannica \cite{britanica}  by the accuracy of its scientific entries \cite{giles} and overcoming the later by the enormous quantity of available information. A detailed analysis of strong and weak features of Wikipedia is given in \cite{reagle,finn}. 
Unique to Wikipedia is that articles make citations to each other, providing a direct relationship 
between webpages and topics. As such, Wikipedia generates a large directed network of article titles with a rather clear meaning. For these reasons, it is interesting to apply algorithms developed for search engines of World Wide Web (WWW), those like the PageRank algorithm \cite{brin}(see also \cite{meyer,rmp2015}), to analyze the ranking properties and relations between Wikipedia articles. For various language editions of Wikipedia it was shown that the PageRank vector produces a reliable ranking of historical figures over 35 centuries of human history \cite{wikizzs,wikievol,eomwiki9,eomwiki24} and a solid Wikipedia ranking of world universities (WRWU) \cite{wikizzs,lageswiki,rokach}. It has been shown that the Wikipedia ranking 
of historical figures is in a good agreement with the well-known Hart ranking \cite{hart}, 
while the WRWU is in a good agreement with the Shanghai Academic ranking of world universities \cite{shanghai}. 

At present, directed networks of real systems can be very large (about  $4.2$ million articles for the English Wikipedia edition in 2013 \cite{rmp2015} or $3.5$ billion web pages for a publicly accessible web crawl that was gathered by the Common Crawl Foundation in 2012
\cite{vigna}). For some studies, one might be interested only in the particular interactions between a very small subset of nodes compared to the full network size. For instance, in this paper, 
we are interested in capturing the interactions of nodes using the networks extracted from 7 Wikipedia language editions (FrWiki, EnWiki, DeWiki, ItWiki, EsWiki, NlWiki and RuWiki).
We use the network datasets of Wikipedia 2013 described in \cite{eomwiki24}.   
The selected nodes (their Wikipedia articles) are embedded in a huge complex directed network with millions of nodes. Thus, the interactions between these selected sets of nodes should be correctly determined taking into account that there are many indirect links between the webpages via all other nodes of the network. In previous works, a solution to this general problem has been proposed 
in \cite{greduced,politwiki,geop} by defining the reduced Google matrix theory. 
Main elements of reduced Google matrix $\GR$ will be presented in Section~\ref{sec:reduced}.
This approach develops the ideas of scattering theory of nuclear and mesoscopic physics and quantum chaos
adapted to Markov chains and Google matrix \cite{greduced,politwiki}.

%%%%%%%%%%%%%%%%%%%%%%%%%%%%%%%%%%%%%%%%%%%%%%%%%%%%%
\begin{figure*}[ht]
\begin{center}
\includegraphics[scale=0.25]{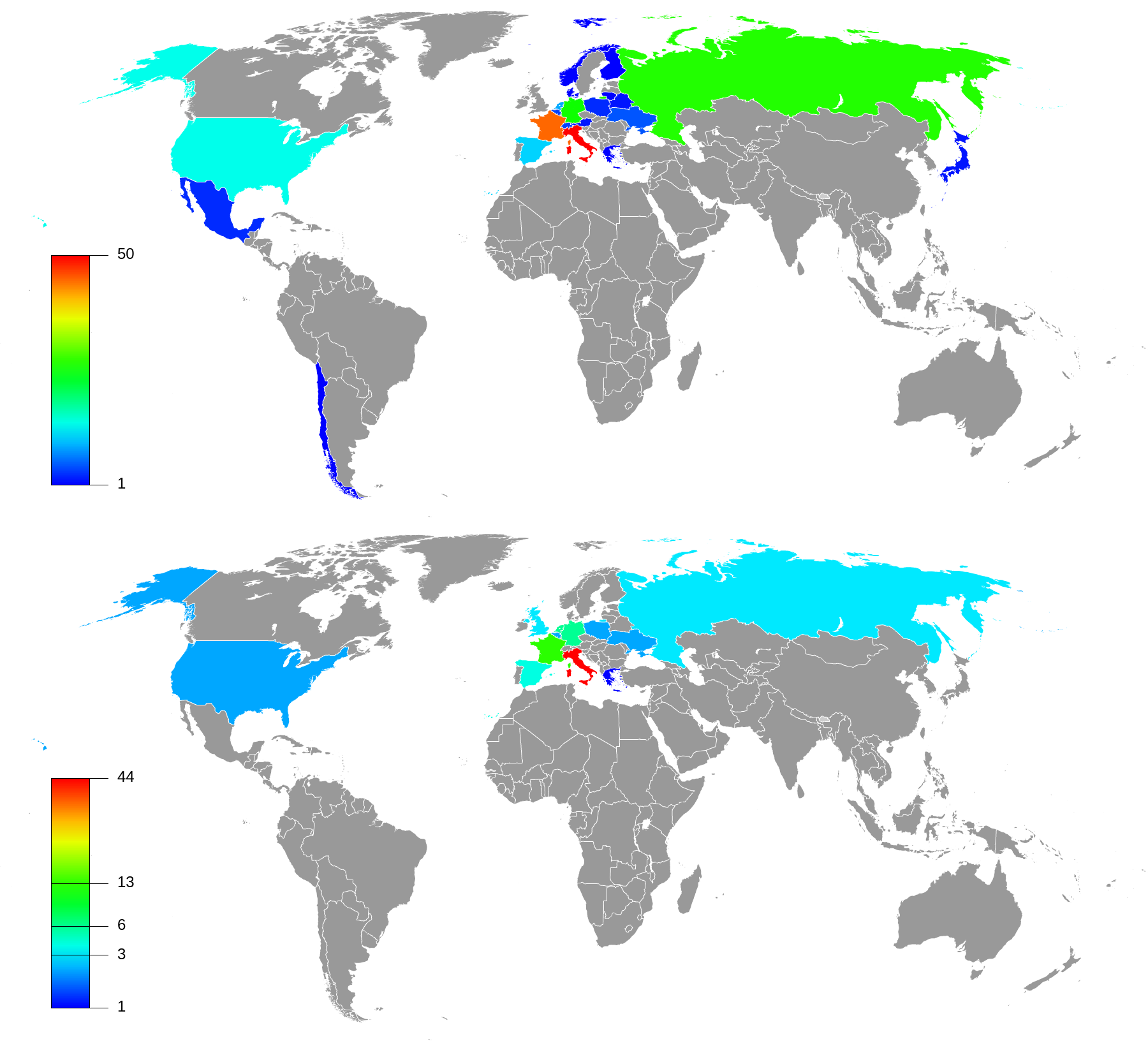}
\caption{{\bf Geographic birthplace distribution of the 223 painters} that 
appear at least one time in the PageRank top 100 painters of one of 
7 language editions analyzed. 
Top panel represents 223 painters for all centuries till present, 
while bottom panel represents 88 painters having middle-age year 
less than year 1800; countries in gray have zero painters.
The birth place is attributed to country borders of 2013.}
\label{fig:birthPlaces}
%fig1paper
\end{center}
\end{figure*}
%%%%%%%%%%%%%%%%%%%%%%%%%%%%%%%%%%%%%%%%%%%%%%%%%%%%%

In a few words, $\GR$  captures in a $N_r$-by-$N_r$\footnote{$N_r$ represents the number of our selected nodes of interest.} Perron-Frobenius matrix the full contribution of both direct and indirect interactions existing in the regular Google matrix model of the network, but only for the reduced set of $N_r$ nodes. The number $N_r$ is in the order of a few tens of nodes, which is considerably smaller that the size of the full Wikipedia network which contains  millions of nodes. Elements of reduced matrix $\GR(i,j)$ can be interpreted as the probability for a random surfer starting at webpage $j$ to arrive in webpage $i$ using direct and indirect interactions. Indirect interactions refer to paths composed in part of webpages different from the $N_r$ ones of interest. Even more interesting and unique to reduced Google matrix theory, 
we show here that intermediate computation steps of $\GR$ offer a decomposition of $\GR$ into matrices that clearly distinguish direct from indirect interactions. 
As such, it is possible to extract a meaningful probability for an indirect interaction 
between two nodes to happen as shown in the results of \cite{politwiki,geop}. 
Thus the reduced Google matrix theory is a perfect candidate for analyzing 
the direct and indirect interactions between the selected painters. 

In this paper, we extract from $\GR$ and its decomposition into direct and indirect matrices 
a high-level \emph{reduced network of $N_r$ painters}. This high-level network is computed 
with both direct and  hidden (i.~e.~ indirect) interactions. More specifically, we deduce from $\GR$ 
a fine-grained classification of painters that captures what we call the \emph{hidden friends} of a given node.  
The structure of these graphs provides relevant information that offers new information compared 
to the direct network of relationships.

The aforementioned networks of direct and hidden interactions can be calculated 
for different Wikipedia language editions. In this paper, reduced Google matrix analysis 
is applied to the set of 30 painters and the set of 40 painters with 40 countries, 
from seven different Wikipedia language editions 
(English, French, German, Spanish, Russian, Italian and Dutch). 
We will refer to these editions using 
EnWiki, FrWiki, DeWiki, EsWiki, RuWiki, ItWiki and NlWiki in the remainder of this paper\footnote{The networks of  EnWiki, FrWiki, RuWiki, DeWiki, ItWiki, EsWiki and NlWiki  contain 4.212, 1.353 , 0.966, 1.533, 1.017, 0.974  and 1.14  millions of articles respectively.}. 
In total we analyzed the list of 3249  painters taken from \cite{allnames}, restricted to the ones that are present in all 7 language editions. Moreover, we provide hereafter, an analysis of the influence of top PageRank painters on world countries after constructing the reduced Google matrix composed of top 40 PageRank painters and the top 40 PageRank countries investigated in \cite{geop}.
We present the full lists of painters, rank lists and additional figures at \cite{ourwebpage}.

This paper introduces first the main elements of reduced Google matrix theory in Section~\ref{sec:reduced}.
Next, Section~\ref{sec:painters} presents the ranking and selection of painters based on the PageRank
algorithm. In Section~\ref{sec:matrices}
the reduced Google matrices are calculated and described for selected sets 
for seven different language editions. Specific emphasis is given to the very different 
English, French and German editions. Then, networks of friendship from direct and  
hidden interaction matrices are created and discussed in Section~\ref{sec:friends}.  
We show that the networks of friends completely 
capture the well-established history of painting by $i)$ interconnecting densely painters of 
the same movement and $ii)$ showing reasonable links between painters of different movements. 
We also obtain the global ranking of painters
averaged over all 7 Wikipedia editions and analyze the interactions between them.
The influence of painters on world countries is analyzed in Section 
\ref{sec:paintersSensitivity}.  
Finally, Section \ref{sec:conclu} discusses featured results and concludes this paper.

%%%%%%%%%%%%%%%%%%%%%%%%%%%%%%%%%%%%%%%%%%
\begin{table*}[ht]
\centering
\caption{Top 40 painters ranked by decreasing importance following $\Theta_P$-score computed over 7 editions. The average PageRank $K_{av}$ is given as well. It derives from $\GR_{av}$, the matrix average of the individual $\GR$ of all 7 editions. }
\label{tab:painters40}
\begin{tabular}{|c|c|c|c|c|c|}
\hline
\textbf{$\Theta_P$ rank} & \textbf{$K_{av}$ rank} & \textbf{Painter} & \textbf{$\Theta_P$ rank} & \textbf{$K_{av}$ rank} & \textbf{Painter} \\ \hline
1                      & 1                    & Vinci            & 21                     & 18                   & Bondone          \\ \hline
2                      & 2                    & Picasso          & 22                     & 25                   & Kandinsky        \\ \hline
3                      & 6                    & Van Gogh             & 23                     & 19                   & Botticelli       \\ \hline
4                      & 4                    & Rijn             & 24                     & 21                   & Caravaggio       \\ \hline
5                      & 5                    & Rubens           & 25                     & 23                   & Velázquez        \\ \hline
6                      & 8                    & Durer            & 26                     & 30                   & Degas            \\ \hline
7                      & 9                    & Titian           & 27                     & 26                   & Bruegel Eld      \\ \hline
8                      & 11                   & Monet            & 28                     & 29                   & Dyck             \\ \hline
9                      & 12                   & Dali             & 29                     & 28                   & Renoir           \\ \hline
10                     & 14                   & Cézanne          & 30                     & 31                   & Chagall          \\ \hline
11                     & 3                    & Michelangelo     & 31                     & 33                   & Lautrec          \\ \hline
12                     & 7                    & Raphael          & 32                     & 27                   & Vermeer          \\ \hline
13                     & 10                   & Goya             & 33                     & 36                   & Poussin          \\ \hline
14                     & 13                   & Vasari           & 34                     & 37                   & Turner           \\ \hline
15                     & 16                   & Matisse          & 35                     & 38                   & Braque           \\ \hline
16                     & 15                   & Warhol           & 36                     & 32                   & Blake            \\ \hline
17                     & 17                   & Delacroix        & 37                     & 34                   & Greco            \\ \hline
18                     & 22                   & Manet            & 38                     & 39                   & Miró             \\ \hline
19                     & 20                   & David            & 39                     & 35                   & Munch            \\ \hline
20                     & 24                   & Gauguin          & 40                     & 40                   & Eyck             \\ \hline
\end{tabular}
\end{table*}
%%%%%%%%%%%%%%%%%%%%%%%%%%%%%%%%%%%%%%%%%%

\section{Reduced Google matrix theory}\label{sec:reduced}

It is convenient to describe the network of $N$ 
Wikipedia articles by the Google matrix $G$ constructed from 
the adjacency matrix $A_{ij}$ with elements $1$ if article (node) $j$ 
points to  article (node) $i$ and zero otherwise. 
Elements of the Google matrix take the standard form 
$G_{ij} = \alpha S_{ij} + (1-\alpha) / N$ \cite{brin,meyer,rmp2015},
where $S$ is the matrix of Markov transitions with elements  $S_{ij}=A_{ij}/k_{out}(j)$, 
$k_{out}(j)=\sum_{i=1}^{N}A_{ij}\neq0$ being the node $j$ out-degree
(number of outgoing links) and with $S_{ij}=1/N$ if $j$ has no outgoing links (dangling node). 
The quantity $0< \alpha <1$ is the damping factor which for a random surfer
determines the probability $(1-\alpha)$ to jump to any node; below we use the standard value $\alpha=0.85$. 
The right eigenvector
of $G$ with the unit eigenvalue gives the PageRank probabilities
$P(j)$ to find a random surfer on a node $j$. We order nodes by decreasing probability $P$ 
getting them ordered by the PageRank index $K=1,2,...N$ with a maximal probability at $K=1$.
From this global ranking we capture the top 50 painters mentioned in 
Tab.~\ref{tab:tableall} for 7 editions.

%%%%%%%%%%%%%%%%%%%%%%%%%%%%%%%%%%%%%%%%%%%%%%%%%%%%%
\begin{table*}[ht]
\begin{center}
\caption{List of names of 30 selected painters in the \emph{Painting categories network} set. They are grouped by categories, and in each category they are ranked following the $\Theta_P$-score obtained for FrWiki, EnWiki and DeWiki. Local PageRank order for FrWiki, EnWiki and DeWiki are given as well. A color is assigned to each category.  
%Red, Blue, Green, Orange and Pink represents Cubism, Fauvism, Impressionists, Great masters and Modern(20-21) respectively
}
\begin{tabular}{|c|c|c|c|c|c|}
\hline
Name                  & Category                  & Colour & FrWiki & EnWiki & DeWiki \\ \hline
Picasso         & Cubism                    & Red    & 1      & 2      & 2      \\ \hline
Braque        & Cubism                    & Red    & 17     & 20     & 20     \\ \hline
L\'{e}ger         & Cubism                    & Red    & 19     & 24     & 24     \\ \hline
Mondrian         & Cubism                    & Red    & 25     & 22     & 22     \\ \hline
Gris             & Cubism                    & Red    & 29     & 28     & 25     \\ \hline
Delaunay       & Cubism                    & Red    & 28     & 27     & 26     \\ \hline
Matisse         & Fauvism                   & Blue   & 6      & 11     & 12     \\ \hline
Gauguin          & Fauvism                   & Blue   & 13     & 15     & 18     \\ \hline
Derain          & Fauvism                   & Blue   & 22     & 25     & 27     \\ \hline
Dufy            & Fauvism                   & Blue   & 27     & 26     & 29     \\ \hline
Rouault       & Fauvism                   & Blue   & 30     & 30     & 28     \\ \hline
Vlaminck   & Fauvism                   & Blue   & 24     & 29     & 30     \\ \hline
Monet          & Impressionists            & Green  & 4      & 9      & 11     \\ \hline
C\'{e}zanne          & Impressionists            & Green  & 8      & 12     & 9      \\ \hline
Manet         & Impressionists            & Green  & 12     & 13     & 16     \\ \hline
Renoir & Impressionists            & Green  & 15     & 14     & 17     \\ \hline
Degas           & Impressionists            & Green  & 18     & 16     & 21     \\ \hline
Pissarro      & Impressionists            & Green  & 23     & 19     & 23     \\ \hline
da Vinci     & Great masters & Orange & 2      & 1      & 1      \\ \hline
Michelangelo          & Great masters & Orange & 3      & 3      & 4      \\ \hline
Raphael               & Great masters & Orange & 5      & 4      & 5      \\ \hline
Rembrandt    & Great masters & Orange & 9      & 5      & 6      \\ \hline
Rubens     & Great masters & Orange & 10     & 7      & 7      \\ \hline
Durer        & Great masters & Orange & 14     & 8      & 3      \\ \hline
Dali         & Modern 20-21              & Pink   & 7      & 10     & 13     \\ \hline
Warhol           & Modern 20-21              & Pink   & 11     & 6      & 8      \\ \hline
Kandinsky     & Modern 20-21              & Pink   & 20     & 17     & 10     \\ \hline
Chagall          & Modern 20-21              & Pink   & 21     & 18     & 15     \\ \hline
Mir\'{o}             & Modern 20-21              & Pink   & 16     & 21     & 19     \\ \hline
Munch          & Modern 20-21              & Pink   & 26     & 23     & 14     \\ \hline
\end{tabular}
\label{tab:painters}
\end{center}
\end{table*}
%%%%%%%%%%%%%%%%%%%%%%%%%%%%%%%%%%%%%%%%%%%%%%%%%%%%%

Reduced Google matrix is constructed for a selected subset of
nodes (articles) following the method described
in \cite{greduced,politwiki,geop} 
and based on concepts of scattering theory 
used in different fields including mesoscopic and nuclear physics, and  
quantum chaos. It captures in a $N_r$-by-$N_r$ Perron-Frobenius matrix 
the full contribution of direct and indirect interactions happening 
in the full Google matrix between the $N_r$ nodes of interest. In addition the 
PageRank probabilities of selected $N_r$ nodes are the same 
as for the global network with $N$ nodes,
up to a constant multiplicative factor taking into account that 
the sum of PageRank probabilities over $N_r$
nodes is unity. Elements of reduced matrix $\GR(i,j)$ 
can be interpreted as the probability for a random surfer starting at 
web-page $j$ to arrive in web-page $i$ using direct and indirect interactions. 
Indirect interactions refer to paths composed in part of web-pages different 
from the $N_r$ ones of interest.    
Even more interesting and unique to reduced Google matrix theory, 
we show here that intermediate computation steps of $\GR$ offer 
a decomposition of $\GR$ into matrices that clearly distinguish 
direct from indirect interactions: $\GR = \Grr + \Gpr + \Gqr$ \cite{politwiki}.
Here $\Grr$ is given by the direct links between selected 
$N_r$ nodes in the global $G$ matrix with $N$ nodes, 
 $\Gpr$ is rather close to 
the matrix in which each column is given by 
the PageRank vector $P_r$, ensuring that PageRank probabilities of $\GR$ are 
the same as for $G$ (up to a constant multiplier).
Therefore $\Gpr$ doesn't provide much information about direct 
and indirect links between selected nodes.
The one playing an interesting role is $\Gqr$, which takes 
into account all indirect links between
selected nodes appearing due to multiple paths via 
the global network nodes $N$ (see~\cite{greduced,politwiki,geop}).
The matrix  $\Gqr = \Gqrd + \Gqrnd$ has diagonal ($\Gqrd$)
and non-diagonal ($\Gqrnd$) parts. Thus $\Gqrnd$
 describes indirect interactions between nodes.
The matrix elements of $\GR$, $\Grr$, $\Gqrnd$ are represented 
in a two dimensional density plot in Fig.~\ref{fig:EnWiki}
for a group of 30 painters of EnWiki. 
The explicit formulas as well as the mathematical and numerical computation 
methods of all three components of $\GR$ are given 
in \cite{greduced,politwiki,geop}. 
We discuss the properties of these matrix components below, 
but before that we introduce our painter selection process for the seven Wikipedia editions.

\section{Selection of painters}\label{sec:painters}
\subsection{Top PageRank painters}
We are interested in this part in selecting the most influential painters representative of the seven investigated Wikipedia editions. 
Importance of nodes is measured in this selection process with the PageRank centrality. 

A Matlab script has been written to retrieve all the painters' names from the ``List of painters by name'' 
webpage~\cite{allnames} edited by Wikipedia that lists painters from all ages and various parts of the world. 
We have collected {\it 3334} names. 
Next, we get their nodes' number in our network representation of each Wikipedia edition. 
Note that some names are not necessarily known for their painting art production (e.g. Hitler). 
Thus we have made a second check to remove such cases from our list of painters. 
This initial sorting leads to a group of 3249 distinct names for our  7 selected editions enlisted in \cite{ourwebpage}.

A Google matrix is constructed for each Wikipedia edition following 
the standard rules described in Section \ref{sec:reduced}. From the Google matrix
of a given edition, PageRank index $K$ of all $N$ nodes is determined. 
From this vector of $N$ values, we extract PageRank nodes of identified painters
and we reorder them by decreasing PageRank value getting local PageRank index of painters. 
Tab.~\ref{tab:tableall} shows the list of the top 50 PageRank painters captured individually by the 7 
selected Wikipedia editions. 
Not surprisingly, the order of top painters changes with respect to 
the editions due to cultural bias but there are some main trends, e.g.:
\begin{itemize}
\item [-] Leonardo Da Vinci ranks first place in 5 out of 7 editions, 
\item [-] Michelangelo and Picasso belong to the top 4 in all editions,  
\item [-] Russian painters, like Viktor Vasnetsov and Ivan Aivazovsky, 
are in the top 20 of RuWiki but don't appear before rank 50 in other editions. 
%\item [-] The German painter Durer is third in DeWiki, but 19th in FrWiki.
\end{itemize}

Using the PageRank of all 3249 painters computed for 7 language editions, 
we have extracted 223 painters by creating the union set of 
top 100 painters of each language edition. The top panel of Fig.~\ref{fig:birthPlaces} 
illustrates the statistics of birth countries for these 223 painters
 (country borders are taken for year 2013). 
There is a clear predominance of European painters in this selection 
with a strong part of Russian artists as well.  
Among these 223 painters, 88 were born before 
year 1800 and the distribution of these 88 painters over the world map demonstrates
a clear dominance of Italy  at these times as to be seen 
in the bottom panel of Fig.~\ref{fig:birthPlaces}.

\subsection{Global ranking of painters}  
\label{sec:global}

The above results demonstrate different cultural views on the importance of painters 
in the different language edition of Wikipedia. To get a global, multi-cultural importance of painters we
use the approach proposed in \cite{eomwiki9,eomwiki24}. It defines a global rank $\Theta_{P}$ with:
\begin{equation}
\Theta_{P} = \sum_{E} (101-R_{P,E}) . 
\label{theta} 
\end{equation}
Here $R_{P,E}$ is the rank of the top 100 painters 
$P$ in Wikipedia edition $E$ retrieved with PageRank algorithm.
The painters with the largest $\Theta_P-$score are the most important ones for $E$ Wikipedia editions.
Based on $\Theta_{P}$ score, we have selected two different sets of painters for our investigations.

\paragraph{Top 40 painters}

From the $\Theta_P-$score calculated for the $E=7$ Wikipedia editions of interest 
(e.g.~EnWiki, FrWiki, RuWiki, DeWiki, ItWiki, EsWiki and NlWiki), we have selected 
the top 40 painters enlisted in Tab.~\ref{tab:painters40} by order of importance.

%%%%%%%%%%%%%%%%%%%%%%%%%%%%%%%%%%%%%%%%%%%%%%%%%%%%%
\begin{figure*}[bth]
\begin{center}
\includegraphics[scale=0.21]{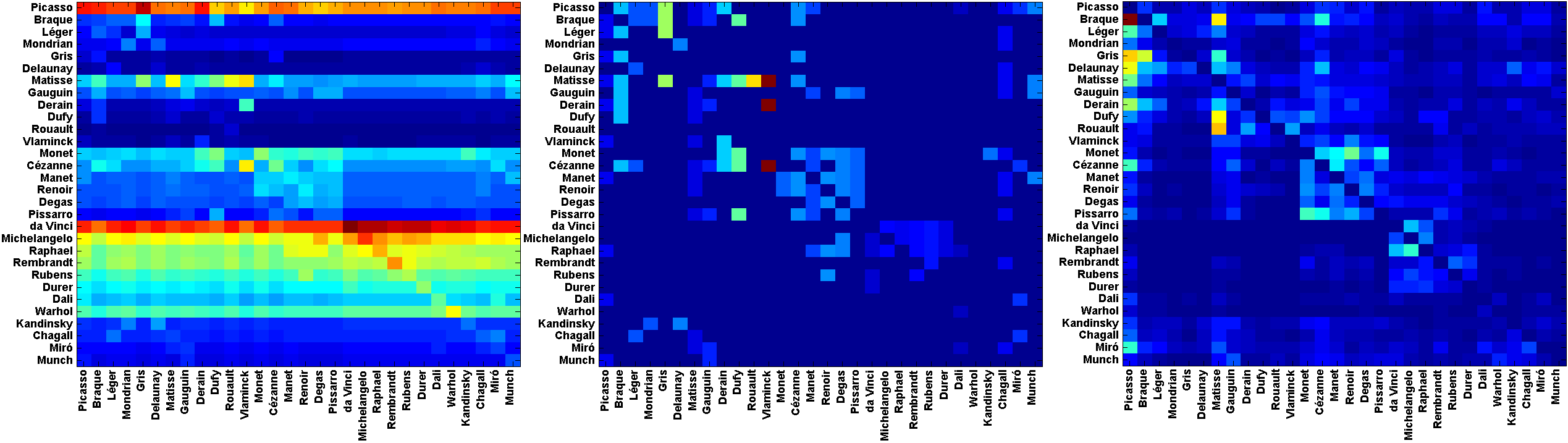}
\caption{{\bf Density plots} of $\GR$ (left), 
$\Grr$ (middle) and $\Gqrnd$ (right) for the reduced network of 30 painters grouped by categories
from Tab.~\ref{tab:painters} for EnWiki network. 
%and the network of 40 painters and 40 countries (bottom)
Color scale represents maximum values in red, intermediate in green and minimum 
close to zero in blue.}
\label{fig:EnWiki}
%fig2paper
\end{center}
\end{figure*}
%%%%%%%%%%%%%%%%%%%%%%%%%%%%%%%%%%%%%%%%%%%%%%%%%%%%%

\paragraph{Painting categories network}

This second set has been chosen to illustrate the existence of 
painting movements and how the reduced Google matrix analysis captures them automatically.
This set is composed of 30 painters that belong to the six following painting categories: Cubism, 
Impressionism, Fauvism, Great masters and Modern art (20th century). 

Following an average ranking $\Theta_P-$score calculated for 3 Wikipedia editions (EnWiki, FrWiki and DeWiki), 
we have selected the top 5 painters of each category. They are enlisted in Tab.~\ref{tab:painters} 
by order of appearance. 
This Table also lists local PageRank index for painters in the French, English and German Wikipedia editions. 
Painters that belong to the same movement or having a common piece of history may probably exhibit 
stronger interactions in Wikipedia. As such, we have created a color code that groups 
together painters that either belong to the same movement (e.g. Fauvism, Cubism, Impressionists) or 
share a big part of history (e.g. Great Masters, Modern). 
Color code is as follows: Red, Blue, Green, Orange and Pink 
represents Cubism, Fauvism, Impressionists, Great masters 
and Modern (20-21st century), respectively.

\section{Reduced Google matrices}\label{sec:matrices}

To illustrate the matrices derived by the reduced Google matrix analysis, 
we plot $\GR$, $\Grr$ and $\Gqrnd$ in Fig.~\ref{fig:EnWiki} for the set of 
30 painters composing the \emph{Painting categories network} in Tab~\ref{tab:painters}. 
The targeted edition is EnWiki.

Columns and lines are ordered with the order set in Tab~\ref{tab:painters}. Following observations can be made.  
$\GR$ is per-column normalized and dominated by the projector $\Gpr$ contribution, 
which is proportional to the global PageRank probabilities 
(for more details see in~\cite{greduced,politwiki}). 
As such, we clearly see that the density of each line of $\GR$ is proportional 
to the importance of the painter in the full network. 
The matrices are interpreted in the following way: painter of column $j$ 
is linked with the probability of element $(i,j)$ to the painter of line $i$.
 
The matrix 
$\Grr$ provides information only on direct links between painters. 
In other words, it represents the probability for a random surfer 
to reach the painter of line $i$ from the article of the painter of column $j$ 
using a hyperlink linking article $j$ to article $i$ in Wikipedia. 
On the contrary, $\Gqrnd$ offers a much more unified view of painters interactions 
as it captures more general indirect (or hidden) interactions via 
the $N-N_r$ other nodes of the full Wikipedia network. In other words, it represents 
the probability linking the painter of column $j$ to the painter of 
line $i$ related to all indirect paths linking article $j$ to article $i$ 
in the full network. An indirect path starts with a hyperlink 
linking the article of painter $j$ to an article $k$ that doesn't belong 
to the $N_r$ nodes and ends with a hyperlink ending on the article of node $i$.    

Reading Fig.~\ref{fig:EnWiki}, we can extract strong and meaningful 
interactions between painters. New links appearing in $\Gqrnd$ 
and being absent from $\Grr$ exist. As an example we list 
the links between Picasso and Braque, Pissaro and Monet, Rouault and Matisse. 
These relationships are very well known in art history, but looking 
at the pure structure of the network (i.~e.~ reading $\Grr$ matrix), they are absent. 
They appear clearly in the higher order mathematical analysis of the network 
using $\Gqrnd$. For instance, it is common knowledge that since his visit 
to Picasso's studio, Braque became impressed by Picasso's paintings. 
They even became friends~\cite{picassobraque}, which confirms our result. 
Pissaro and Monet are both impressionists. Monet succeeded in reaching England 
after entrusting a number of his works to Pissaro~\cite{pissaromonet}. 
Rouault and Matisse were both students of Gustave Moreau~\cite{romatisse1} and 
were deeply influenced by him throughout their life~\cite{romatisse2}. 
Their relationship began in 1906 and lasted all their life. 
All these interactions can be extracted from the network of Wikipedia 
webpages using $\Gqrnd$ matrix. 

In order to simplify the reading and interpretation of these matrices, 
we have introduced in \cite{greduced,politwiki,geop} a set of tools that captures essential features 
of the reduced network. In Section \ref{sec:friends}, we build the friendship networks 
for our sets of painters and in Section~\ref{sec:paintersSensitivity} 
we analyze the influence of painters on countries using the PageRank sensitivity analysis of $\GR$.

%table of friends
%%%%%%%%%%%%%%%%%%%%%%%%%%%%%%%%%%%%%%%%%%%%%%%%%%%%%
%\begin{table*}[ht]
%\centering
%\caption{Cross-editions friends from $\Gqrnd$ for the top painter of each category. 
%For each top painter, we list the friends present in the friends list given by 
%6 Wikipedia editions of Tab.~\ref{tab:tableall}, 
%the ones present in 5 editions out of 6 and the ones present in 4 editions out of 6.}
%\label{tab:friends_followers}
%\begin{tabular}{|c|c|c|c|}
%\hline
%Top Painter & all 6 editions             & 5 out of 6 editions & 4 out of 6 editions      \\ \hline
%Picasso     &Braque - Gris  &                     &                          \\ \hline
%Matisse       &Rouault                       &               &Braque - Dufy                   \\ \hline
%Monet     &Renoir                     &Pissarro              &                          \\ \hline
%da Vinci     &Michelangelo - Raphael                     &Durer                     &Degas   \\ \hline
%Dali     &Mir\'{o}                     &                &                   \\ \hline
%\end{tabular}
%\end{table*}

%%%%%%%%%%%%%%%%%%%%%%%%%%%%%%%%%%%%%%%%%%%%%%%%%%%%%

%%%%%%%%%%%%%%%%%%%%%%%%%%%%%%%%%%%%%%%%%%%%%%%%%%%%%
\begin{figure*}
\begin{center}
\includegraphics[scale=0.25]{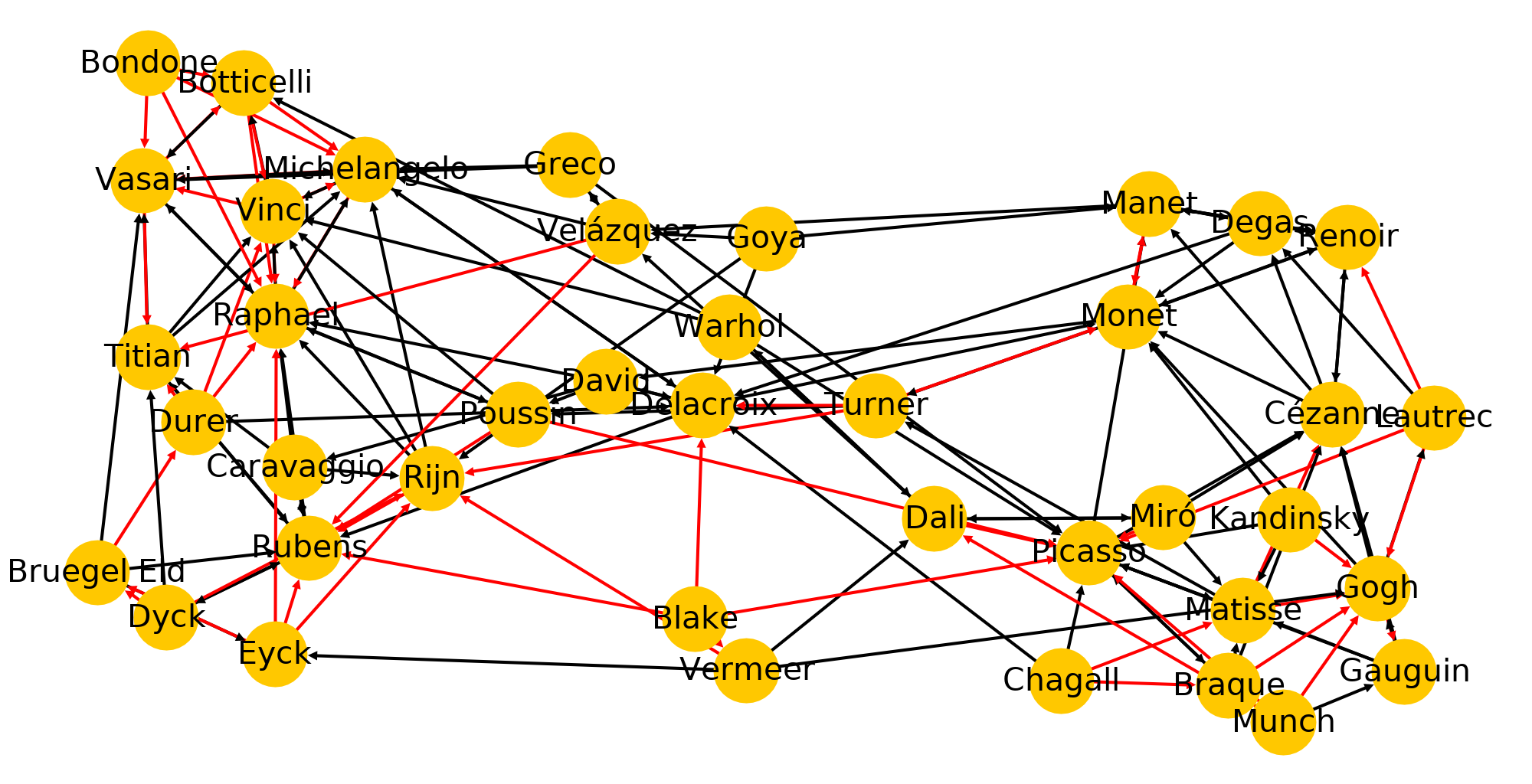}
\includegraphics[scale=0.25]{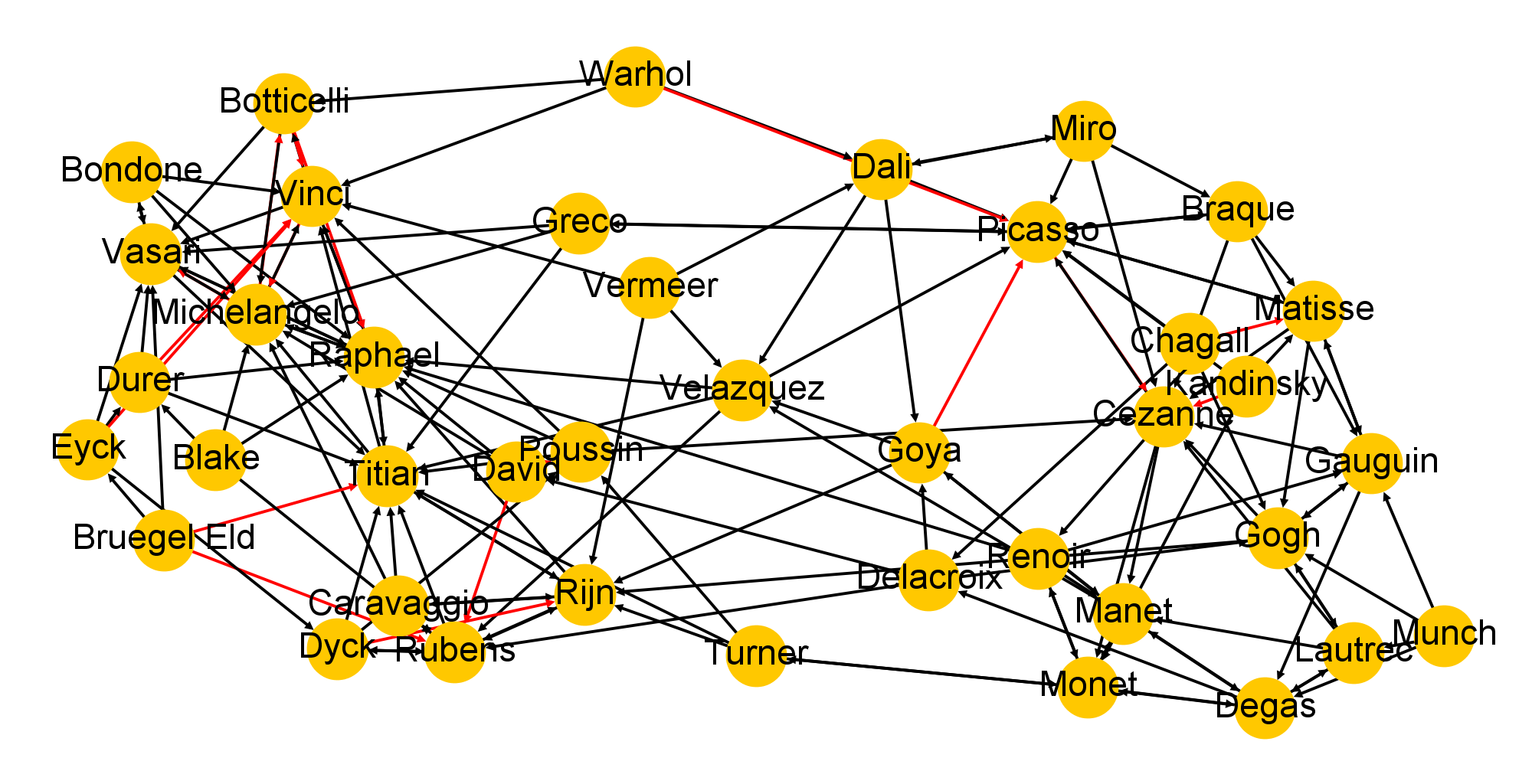}
\caption{{\bf Friendship networks for the top 40 painters data set} listed in Tab.~\ref{tab:painters40}.
Top panel: Friendship network extracted from $\Grr+\Gqrnd$ computed with FrWiki.
Bottom panel: Friendship network extracted from the matrix  
$\Grr_{av}+\Gqrnd_{av}$, where $\GR_{av}$ is the average of the $\GR$ matrices obtained for all 7 Wikipedia editions.  
For each painter in the set, arrows are drawn to its top 4 friends. Arrows are colored in red if component $\Gqrnd$ is larger than component $\Grr$, and in black otherwise.
Both graphs are automatically plotted using Yifan Hu layout with {\it Gephi}\cite{gephi}. 
}
\label{fig:global40}
%fig3paper
\end{center}
\end{figure*}
%%%%%%%%%%%%%%%%%%%%%%%%%%%%%%%%%%%%%%%%%%%%%%%%%%%%%

\section{Friendship networks}\label{sec:friends}

\subsection{Friendship network construction}

It is possible to extract from $\Grr$ and $\Gqrnd$ a network of friendship to conveniently illustrate direct and hidden links in the network, or a combination of both. Direct links are extracted from $\Grr$ while hidden (i.~e.~ indirect) are extracted from $\Gqrnd$. The network of friends is built by considering larger matrix elements in a column $j$ of a given painter as top friends (i.~e.~ there is a high probability to end in node $i$ from node $j$). 
It is true that the word \emph{friend} usually represents a symmetrical relationship. But we have chosen this denomination for its ease of use. Clearly, in this paper, a friend represents a node that is an attractor for the node of interest.
%%%%%%%%%%%%%%%%%%%%%%%%%%%%%%%%%%%%%%%%%%%%%%%%%%%%%
\begin{figure*}[ht]
\begin{center}
\includegraphics[scale=0.3]{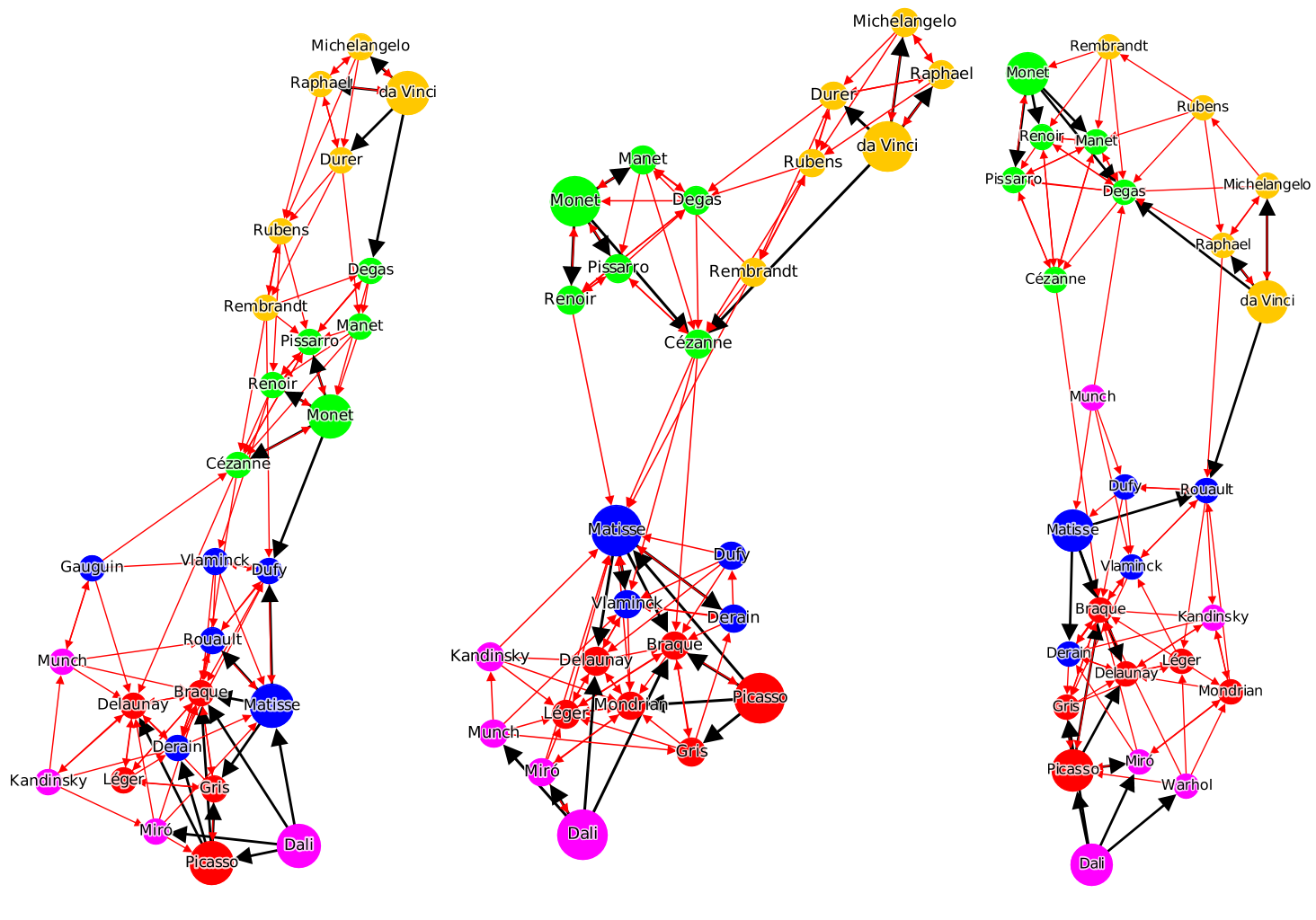}
\caption{{\bf Friendship networks for the painting categories data set} listed of Tab.~\ref{tab:painters}.
Results are extracted from the $\Gqrnd$ matrix derived from EnWiki (left), FrWiki (middle) and DeWiki (right).
Red, Blue, Green, Orange and Pink nodes represents Cubism, Fauvism, Impressionists, 
Great masters and Modern(20-21) respectively.
The top painter node points with a bold black arrow to its top 4 friends. 
Red arrows represent the friends of friends interactions computed until 
no new edges are added to the graph.
All graphs are automatically plotted using Yifan Hu layout with {\it Gephi}\cite{gephi}.
}
\label{fig:networkGqr}
%fig4paper
\end{center}
\end{figure*}
%%%%%%%%%%%%%%%%%%%%%%%%%%%%%%%%%%%%%%%%%%
From the notion of friendship, we derive the networks of friends shown in Fig.~\ref{fig:global40} and Fig.~\ref{fig:networkGqr}. 
In Fig.~\ref{fig:global40}, the set of top 40 painters is analyzed while in Fig.~\ref{fig:networkGqr} the painting categories set is investigated. 
Both networks of friends have been derived in the following way:
\begin{itemize}
\item For Fig.~\ref{fig:global40}, arrows representing 4 outlinks are drawn from each painter to its top 4 friends in the matrix $\Grr+\Gqrnd$. In this figure, we mark arrows in red if the $\Gqrnd$ component (i.~e.~ indirect link probability) is larger than the $\Grr$ component (i.~e.~ direct link probability). Black arrows thus represent the opposite case. The graphs are automatically plotted with {\it Gephi}\cite{gephi} using the Yifan Hu algorithm.
In other words, for each painter $j$ of Tab.~\ref{tab:painters40}, we extract from sum of both matrices the top 4 \emph{Friends} given by the 4 strongest elements of column $j$. This figure represents two different views: $i)$ a regional view (top panel) as $\Grr+\Gqrnd$ are computed for FrWiki and $ii)$ a unified view (bottom panel) as the friendship network is built from $\Grr_{av}+\Gqrnd_{av}$ which is defined as the average of the corresponding matrices computed over all 7 Wikipedia editions. 

\item In Fig.~\ref{fig:networkGqr}, we capture for the Painting categories data set the sole indirect interactions provided by $\Gqrnd$ between the 5 categories of painters. Therefore, we have selected the most influential painter in each category. This category leader is the one with the best (i.~e.~smallest) average ranking score over all 6 selected Wikipedia editions.  
The top painters are Pablo Picasso for Cubism,  Henri Matisse for Fauvism~\cite{schneider}, Claude Monet for Impressionists \cite{wildenstein,montebello},  Leonardo Da Vinci for Great Masters and Dali for Modern. 
The networks of Fig.~\ref{fig:networkGqr} are created by marking with black arrows the link between each leading painter and its top 4 friends in $\Gqrnd$. Red arrows represent the friends of friends interactions computed until no new edges are added to the graph.
Three networks are plotted, originating from EnWiki, FrWiKi and EnWiki.
\end{itemize}

We discuss next the interesting results observed for both networks of painters.

\subsection{Top 40 painters networks}

This part discusses the friendship networks of Fig.~\ref{fig:global40}.
The point of our analysis is to underline the relative importance of direct and indirect interactions. 
For black arrows, the direct component $\Grr$ is stronger than the indirect one $\Gqrnd$. 
For red arrows, the indirect component is the strongest. 

The top panel presents the French Wikipedia view of this painter network. 
A large proportion of arrows are red, meaning that in this case the indirect interaction 
between nodes is contributing a lot to building the network.  
The graph has a clear structure of three main 
clusters: painters who worked in France
(right part around Matisse-Picasso-Monet-Van Gogh),
the ones that have worked in Italy (left top corner
with Da Vinci-Titian-Botticelli)
and the ones that have worked in the Netherlands and Belgium (bottom left corner
with Rubens-Rembrandt Van Rijn-van Dyck).
This shows that our network analysis captures realistic relations between painters.

However, the above presentation takes into account 
only the opinion of the FrWiki edition  
with a dominance of French culture linked to French language.
It is interesting to have the network structure
which takes into account the opinions
of all 7 editions. In fact the approach of the reduced Google matrix 
is well suited for this. Indeed, to perform the average over different cultures 
we take $\GR$ for 40 painters (size 40) and its components
and take the average of these 7 matrices with equal democratic weights getting 
in this way the average $\GR_{av}$ and its average 3 components.
Of course, after averaging $\GR_{av}$ still belongs to the 
class of Google matrices. The network structure obtained
from $\Grr_{av}+\Gqrnd_{av}$ is shown in the bottom panel of Fig.~\ref{fig:global40}.
In global the two centers with painters worked in France (on the right)
and in Italy (left) is similar to the case of FrWiki in the top panel.
However, the number of indirect links (red arrows) is decreased.
We attribute this to increased number of direct links present in all 7 editions.

We also note that $\GR_{av}$ has now a new average PageRank vector
$P_{av}(K_{av})$,
which takes into account opinions of all 7 cultures
(it is different from the simple averaged probabilities of 7 
individual PageRank vectors). This average rank index $K_{av}$
is shown in Tab.~\ref{tab:painters40}. 
The top two positions are the same as for $\Theta$-rank, however, there is a noticeable change of order
in positions 3-12 with more importance given to
ancient Italian masters like Michelangelo, Raphael, Titian 
who moves to the top $K_{av}$ positions
while more recent painters such as Van Gogh, Monet, Dali
are getting larger  $K_{av}$ values. 
We attribute this to the fact that the ancient historical 
figures are on average better reviewed in various cultures
and Wikipedia editions.

%%%%%%%%%%%%%%%%%%%%%%%%%%%%%%%%%%%%%%%%%%%%%%%%%%%%%
\begin{table}[b]
\centering
\caption{List of PageRank of top 40 countries in EnWiki}
\begin{tabular}{|c|c|c|c|}
\hline
\textbf{Order} & \textbf{Country} & \textbf{Order} & \textbf{Country} \\ \hline
1              & US               & 21             & NO               \\ \hline
2              & FR               & 22             & RO               \\ \hline
3              & GB/UK               & 23             & TK               \\ \hline
4              & DE               & 24             & ZA               \\ \hline
5              & CA               & 25             & BE               \\ \hline
6              & IN               & 26             & AT               \\ \hline
7              & AU               & 27             & GR               \\ \hline
8              & IT               & 28             & AR               \\ \hline
9              & JP               & 29             & PH               \\ \hline
10             & CN               & 30             & PT               \\ \hline
11             & RU               & 31             & PK               \\ \hline
12             & ES               & 32             & DK               \\ \hline
13             & PL               & 33             & IL               \\ \hline
14             & NL               & 34             & FI               \\ \hline
15             & IR               & 35             & EG               \\ \hline
16             & BR               & 36             & ID               \\ \hline
17             & SE               & 37             & HU               \\ \hline
18             & NZ               & 38             & TW               \\ \hline
19             & MX               & 39             & KR               \\ \hline
20             & CH               & 40             & UA               \\ \hline
\end{tabular}
\label{tab:countries40}
\end{table}
%%%%%%%%%%%%%%%%%%%%%%%%%%%%%%%%%%%%%%%%%%%%%%%%%%%%%
\subsection{Paintings categories network}

Fig.~\ref{fig:networkGqr} illustrates the indirect interactions provided by 
$\Gqrnd$ in the painting categories network. The 5 leading painters are connected 
with black arrows to the top 4 friends. And these friends are connected to their top 4 friends with red arrows.

Thus $\Gqrnd$ seems to emphasize finer-grained regional interactions and by looking 
at the interactions, we can see the strong relationship between Da Vinci, Michelangelo and Raphael 
which can be explained by the fact that they are the nucleus of fifteenth-century Florentine art~\cite{mich}. 
Another strong relation could be snapped between Mir\`o and Dali, as both are inspired by Picasso~\cite{berger,dali}.

Impressionists, Fauvism, Cubism and Great masters create, in all editions, a cluster of nodes densely interconnected. 
The group of Modern painters plays a role by connecting the other categories: 
\begin{enumerate}
    \item Dali seems to be the common interconnection node between Fauvism and Cubism categories in EnWiki.
    \item Kandinsky connects Fauvism and Cubism in FrWiki.
    \item  Munch connects Impressionists 
and Fauvism in DeWiki. The networks of $\Gqrnd$ end up almost spanning the full set of 30 painters. 
\end{enumerate}

These links show that the interactions between the painters groups are coherent. 
These graphs picture the essence of painting history by grouping together painters that belong 
to the same movement and by interconnecting them in a reasonable and close-to reality way. 

%Impressionists and Great Masters still make a cluster of nodes densely interconnected.
For instance, our graphs are consistent with the history of modern art which starts with 
the Impressionists movement (1870-1890) that searched for the exact analysis of the effects of 
color and light in nature. The painters we have selected are among the most important ones of 
the movement and they create a clear cluster of nodes in Fig.~\ref{fig:networkGqr} (see green nodes) 
as they exhibit a tight relationship in $\GR$. The Fauvism movement emerged after impressionist (1899-1908) \cite{ferrier,brodskaia,whitfield}. Fauvist painters were concerned with the impression created with colors. This movement was inspired by different artists such as Matisse. 
The \emph{Fauves} members were a loosely shaped group of artists with shared interests. 
Henri Matisse became later the leader of the group of artists~\cite{schneider}. He introduced 
unnatural and intense color into their paintings to describe light and space. 
The fauvism movement is the precursor of the Cubism movement~\cite{antliff}.  Our result shows 
deep relationships between Fauvism and Cubism, noting that Braque is always the core of this interconnection. 
Cubism movement (1907- 1922) is pretty distinct from Impressionism, which is underlined as well 
in our graphs with only a few red links connecting these two clusters of nodes.

\section{Influence of painters on countries}
\label{sec:paintersSensitivity}

\subsection{Datasets}
Another complementary study is presented here to visualize the influence of painters on countries. To analyze the relation between painters and countries of the world we construct a reduced Google matrix with $N_r=80$ nodes composed of the top 40 painters shown in Table~\ref{tab:painters40} and the group of 40 countries listed in Table~\ref{tab:countries40}. The painters are the ones having top  $\Theta_P-$score for $E=7$: EnWiki, FrWiki, RuWiki, DeWiki, ItWiki, EsWiki and NlWiki. Table~\ref{tab:painters40} only lists short names, however, the full painter names together 
with their $\Theta_P-$score, birth country and life period are available as well in \cite{ourwebpage}.
The top 40 countries of EnWiki are presented in Table~\ref{tab:countries40}.
The names of countries are given by ISO 3166-1 alpha-2 code (see \cite{wikiiso2}).

%%%%%%%%%%%%%%%%%%%%%%%%%%%%%%%%%%%%%%%%%%%%%%%%%%%%%
\begin{figure}[t]
\begin{center}
\includegraphics[scale=0.15]{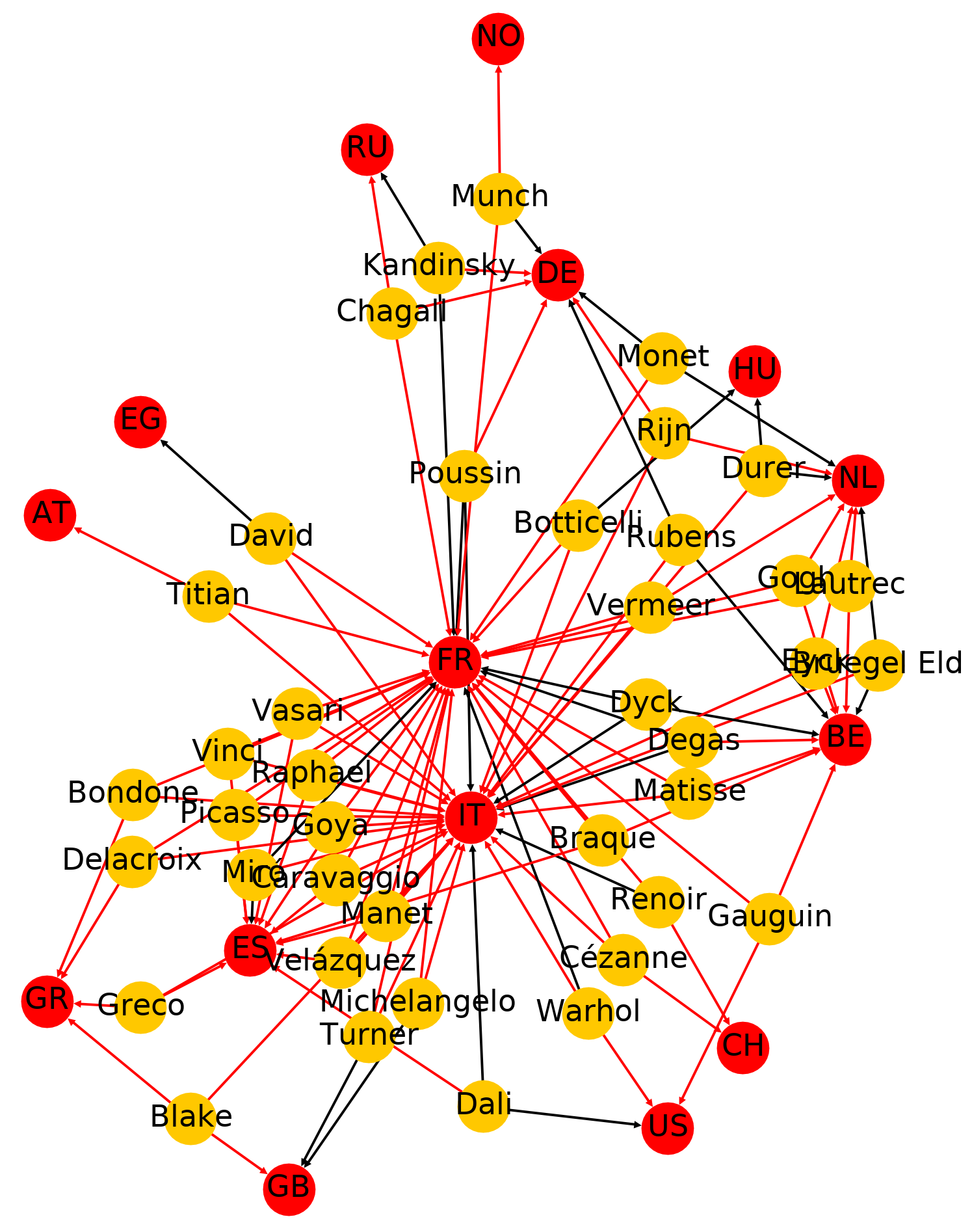}
\end{center}
\caption{{\bf Network structure of top 3 country friends for top 40 painter network for EnWiki}
Painters are selected from the global rank list of 7 Wikipedia editions from Table~\ref{tab:painters40}
for top 40 PageRank countries of EnWiki from Table~\ref{tab:countries40}. 
Arrows are showing links only from a painter to top 3 countries, they are
given by links of matrix elements $\Grr+\Gqrnd$,
red arrow mark links when an element $\Gqrnd$ is larger than element $\Grr$,
black arrows are drown in opposite case. Countries and shown by red circles and painters are shown
by yellow circles.}
\label{fig:3countriesEn}
%fig5paper
\end{figure}

\begin{figure}
\begin{center}
\includegraphics[scale=0.15]{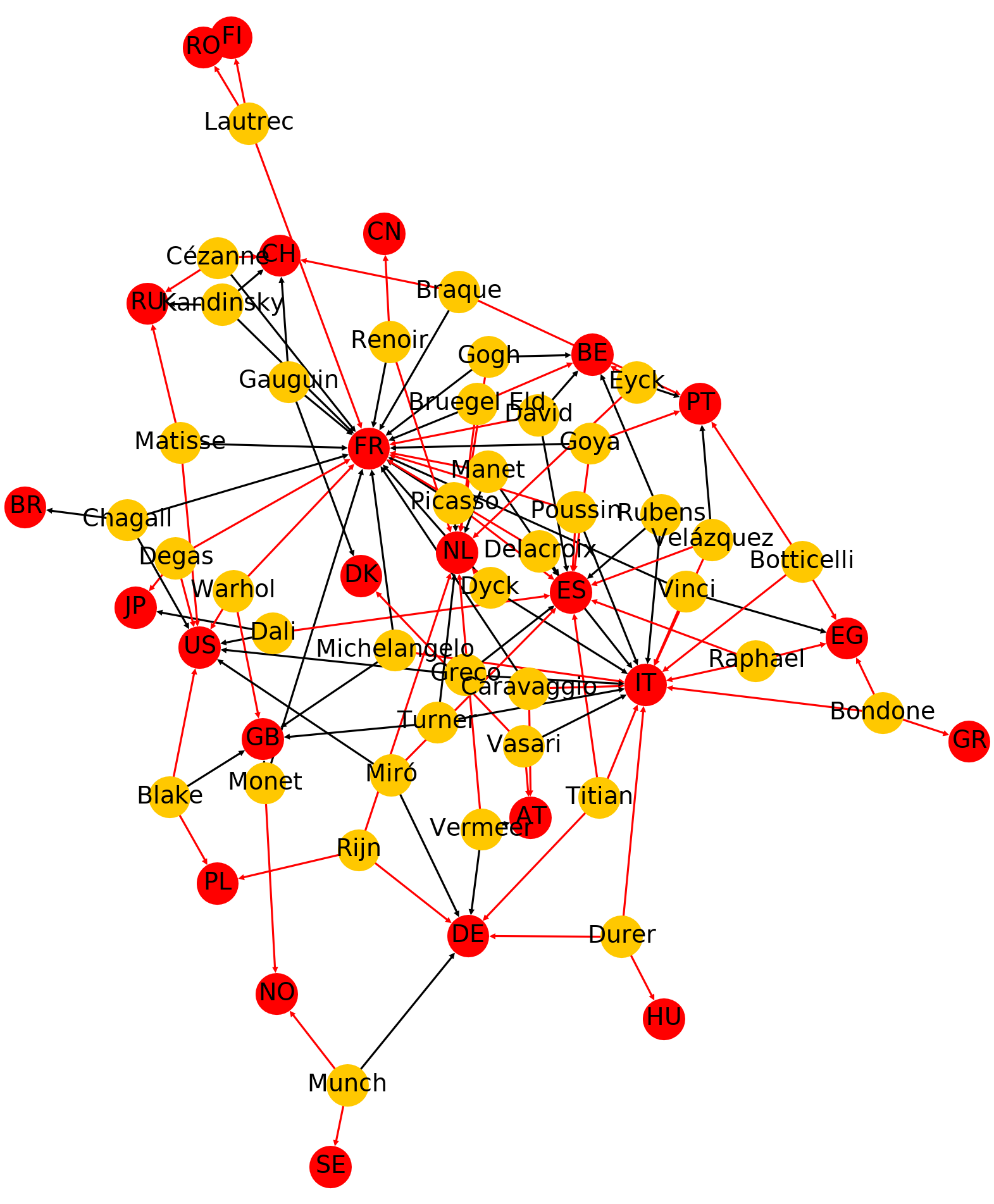}
\end{center}
\caption{{\bf Network structure of top 3 country friends for top 40 painter network for FrWiki}
Painters are selected from the global rank list of 7 Wikipedia editions from Table~\ref{tab:painters40}
for top 40 PageRank countries of EnWiki from Table~\ref{tab:countries40}. 
}
\label{fig:3countriesFr}
%fig6paper
\end{figure}

\begin{figure}
\begin{center}
\includegraphics[scale=0.15]{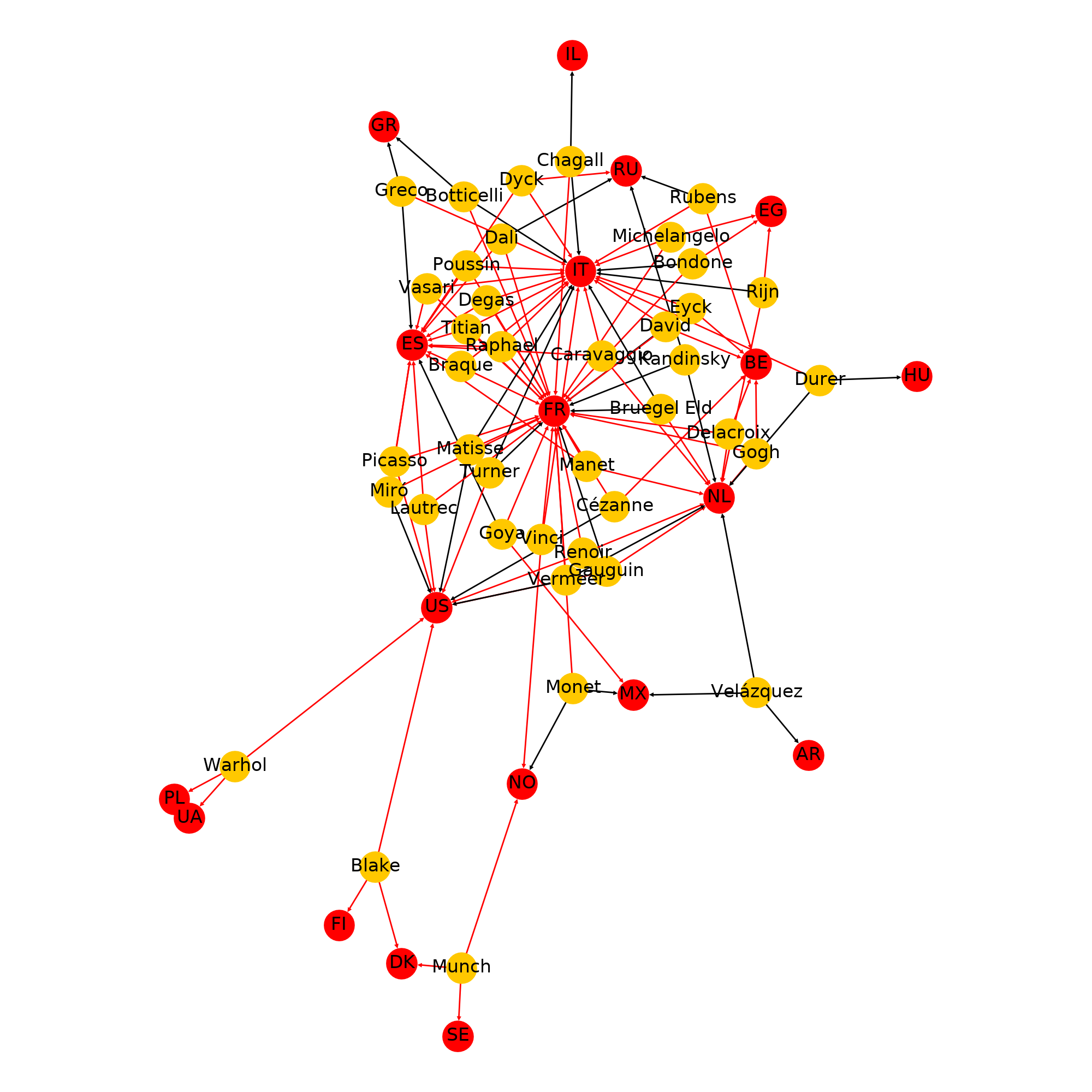}
\end{center}
\caption{{\bf Network structure of top 3 country friends for top 40 painter network for DeWiki}
Painters are selected from the global rank list of 7 Wikipedia editions from Table~\ref{tab:painters40}
for top 40 PageRank countries of EnWiki from Table~\ref{tab:countries40}. 
}
\label{fig:3countriesDe}
%fig7paper
\end{figure}

%%%%%%%%%%%%%%%%%%%%%%%%%%%%%%%%%%%%%%%%%%%%%%%%%%%%%

\subsection{Networks of painters and countries}
The three painters with the largest $\Theta_P-$score are:
\begin{enumerate}
\item Leonardo Da Vinci with $\Theta=698$, born in Italy, 
\item Pablo Picasso with $\Theta=688$, born in Spain,
\item Vincent Van Gogh with $\Theta=656$, born in the Netherlands. 
\end{enumerate}

The following painters are the most important one for their country of birth:
\begin{itemize}
\item Peter Paul Rubens for Germany with $\Theta=651$ (but worked mainly in Netherlands),
\item Claude Monet for France with $\Theta=605$,
\item Wassily Kandinsky for Russia with $\Theta=515$,
\item Joseph Mallord William Turner for United Kingdom (UK or GR) with $\Theta=386$.
\end{itemize}
The top 6 countries with the largest number of painters from the global list of 223 painters
are Italy (50), France (45), Russia (27), Germany (26), USA (14), Spain (11)
(note that the 223 other countries are listed as well in \cite{ourwebpage}).

%%%%%%%%%%%%%%%%%%%%%%%%%%%%%%%%%%%%%%%%%%%%%%%%%%%%%
\begin{figure*}
\begin{center}
\includegraphics[scale=0.25]{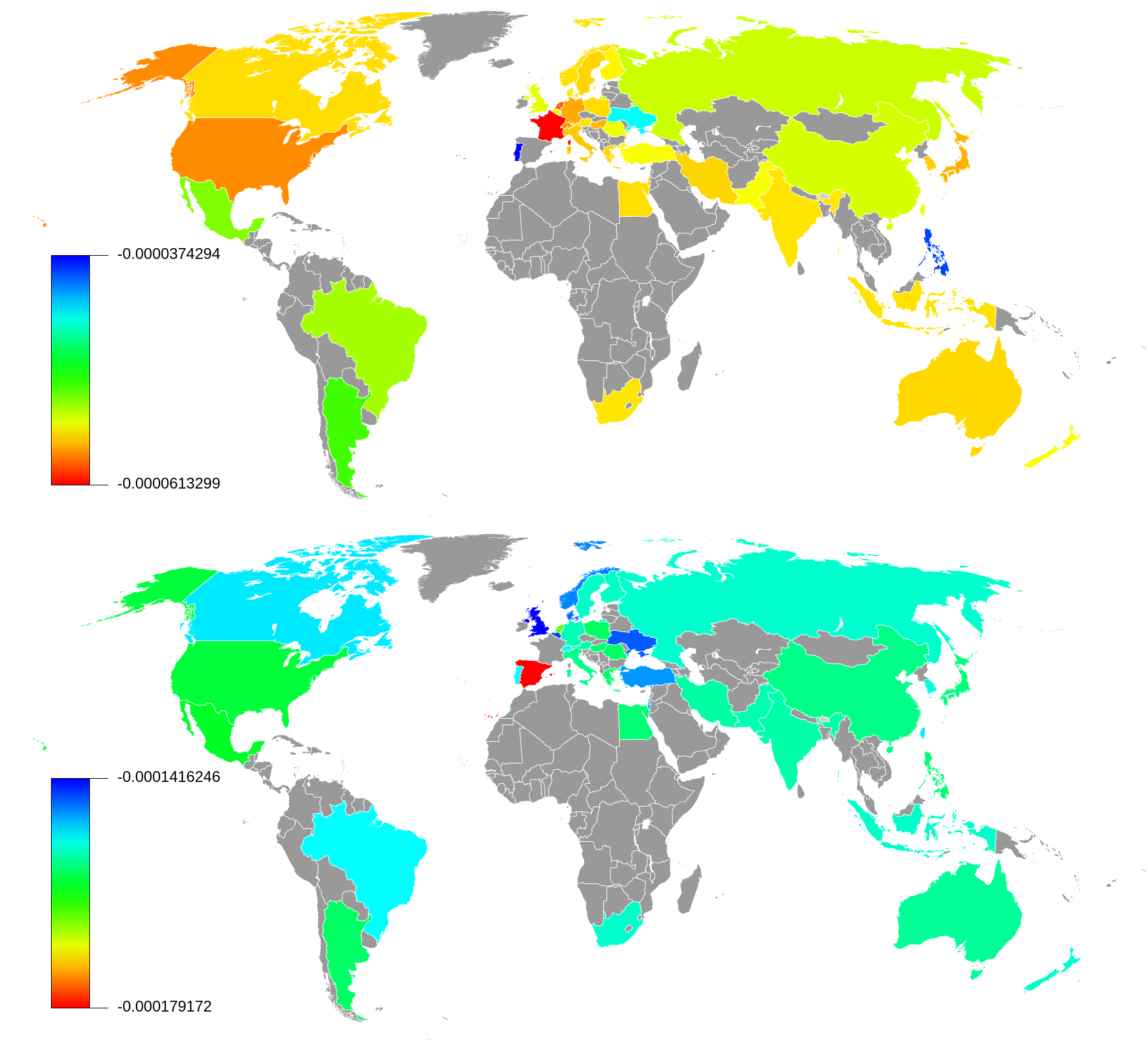}
\end{center}
\caption{{\bf Sensitivity $D$ of 40 world countries to the link variation going from Picasso to Spain and
Picasso to France. 
Top panel: Picasso to Spain and bottom panel: Picasso to France. Data is averaged over 7 Wikipedia editions.
For a better visibility, sensitivity of Spain (top) and France (bottom) are given in Figure~\ref{fig:diagsensi}. 
}}
\label{fig:picassoesfr}
%fig8paper
\end{figure*}

%%%%%%%%%%%%%%%%%%%%%%%%%%%%%%%%%%%%%%%%%%%%%%%%%%%%%

%%%%%%%%%%%%%%%%%%%%%%%%%%%%%%%%%%%%%%%%%%%%%%%%%%%%%
\begin{figure*}
\begin{center}
\includegraphics[scale=0.25]{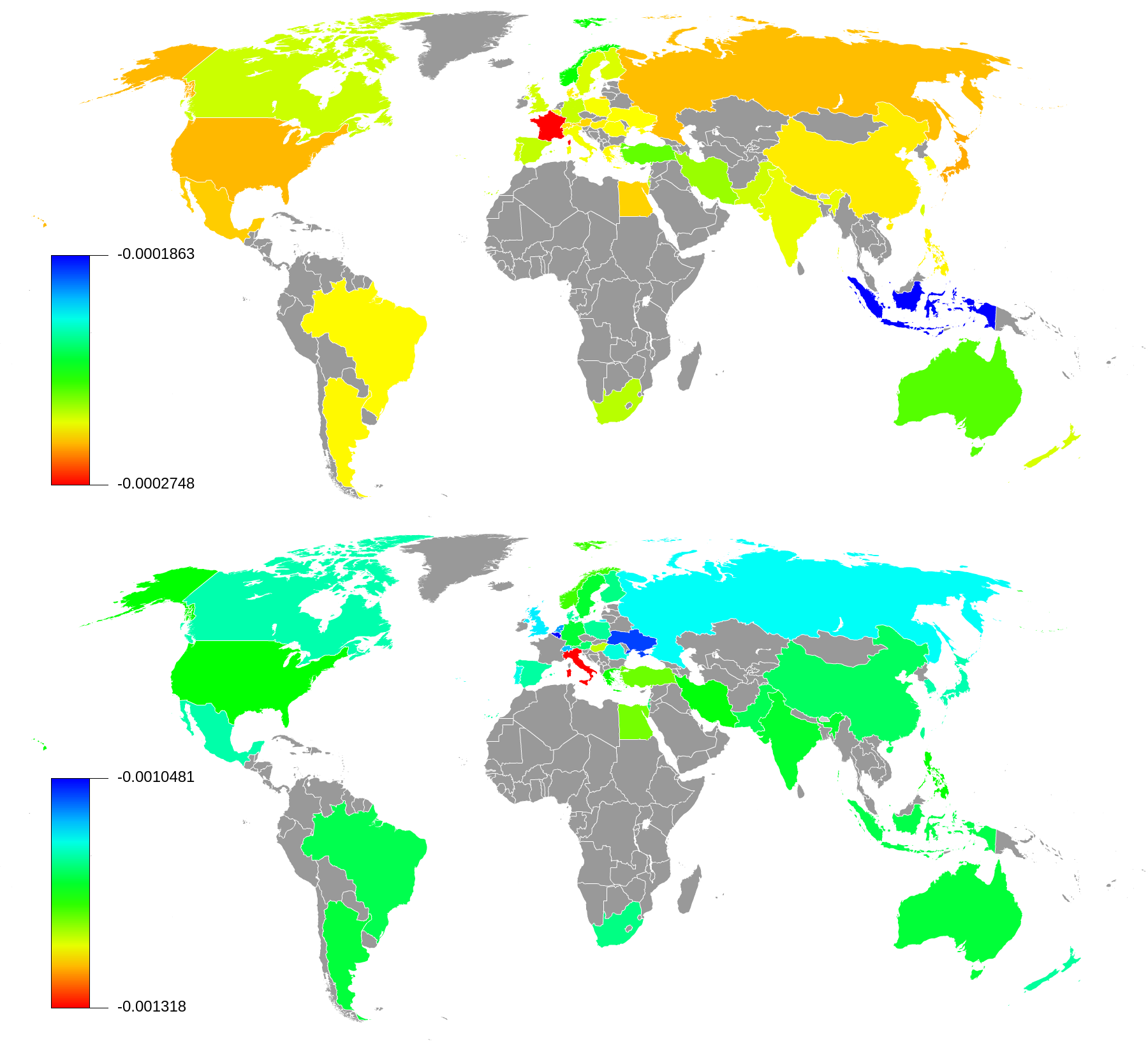}
\end{center}
\caption{{\bf Sensitivity $D$ of 40 world countries
to the link variation going from Van Gogh to the Netherlands and
Da Vinci to France.}
Top panel: Van Gogh-the Netherlands and bottom panel: Da Vinci-France.
Data is averaged over 7 Wikipedia editions.
For a better visibility, sensitivity of the Netherlands (top)
and France (bottom) are given in Figure~\ref{fig:diagsensi}.
}
\label{fig:gohgvinci}
%fig9paper
\end{figure*}
%%%%%%%%%%%%%%%%%%%%%%%%%%%%%%%%%%%%%%%%%%%%%%%%%%%%%

The geopolitical relations between painters and countries has been analyzed more precisely 
for EnWiki, FrWiki and DeWiki data. Therefore, we have plotted a network of friendship between 
our 40 painters and their top 3 most friendly countries. This network has been calculated using 
$\Grr$ and $\Gqrnd$ calculated for the union of 40 painters and 40 countries. 
For each painter column, we select the top 3 countries in the sum matrix $\Grr + \Gqrnd$ 
to account for direct and indirect interactions and mitigate the effect of the projector component. 
Resulting networks are shown in Figure~\ref{fig:3countriesEn}, \ref{fig:3countriesFr} and 
\ref{fig:3countriesDe} for EnWiki, FrWiki and DeWiki, respectively. In these figures, 
arrows are colored in red if $\Gqrnd(i,j) > \Grr(i,j)$ and in black other wise. 
The network structure is different for each edition due to different cultural views and preferences.
However, the central role of France and Italy is well visible in all 3 editions.

\subsection{Influence of painters on countries}  
\label{sec:influence}

To analyze in a more direct way the world influence of painters we average $\GR$ matrix and 
its three components $\Gpr$, $\Grr$, $\Gqr$ over 7 Wikipedia editions that allows us to account 
for different cultural views on selected 40 painters of Table~\ref{tab:painters40} and  
40 countries of Table~\ref{tab:countries40}. 
The  reduced Google matrix $\GR_{av}$ averaged over these 7 different editions allows 
us to obtain a balanced view of various cultural opinions of Wikipedia language editions 
for a selected group of nodes representing Wikipedia articles.
We determine the PageRank probability of this averaged $\GR$ matrix and compute 
its logarithmic derivative (sensitivity) with respect to a weight variation of a selected link 
going from a specific painter to a specific country. For instance, we vary the intensity in 
$\GR$ of the link going from Picasso to Spain, and observe the variation of PageRank for other countries. 
This PageRank probability variation is defined as the sensitivity $D(i)$ of a node $i$ to a link change. 
We refer the reader to \cite{menacomm} for a precise definition of $D$
($D$ essentially is given by a logarithmic derivative of PareRank probability in
respect to a relative link weight variation).

{\it a) Influence of Picasso link to Spain and France:}
Figure~\ref{fig:picassoesfr} shows the sensitivity $D$ of 40 world countries 
with respect to a link variation from Picasso to Spain (top panel)
and from Picasso to France (bottom panel). 
Pablo Picasso, the son of the Spanish painter Don Jos\'{e} Ruiz y Blanco, was born in Spain in 1881.
Pablo began painting since he was eight, and in 1896, he has joined the art 
and design school of Barcelona "Escola de la Llotja". 
In 1904, Picasso married Fernande Olivier a French artist and model. 
Since that, Picasso spent most of his life in France and died there at
92 years old. This could explain the results we have obtained from our sensitivity analysis, 
which underlines that France and Spain are the countries that are mostly affected for 
a Picasso-Spain and a Picasso-France link variation, respectively.

{\it  b) Influence of Van Gogh link to Netherlands and Da Vinci link to France:}
Figure~\ref{fig:gohgvinci} shows the sensitivity $D$ of 40 world countries 
with respect to a link variation from Van Gogh to Netherlands and from da Vinci to France 
in top and bottom panels, respectively.
Even though Van Gogh has only spent the last four years of his life in different places of France, 
these years were important to Van Gogh's painting career. Van Gogh has built there strong relationships 
with leading French painters. He has worked at that time with Emile Bernard, Henri de Toulouse-Lautrec,
 Georges Seurat, Paul Signac and Gauguin. These relationships and the work achieved by Van Gogh 
in France explain our results in the top panel of Figure~\ref{fig:gohgvinci}, 
which shows that France is strongly influenced by a link variation from Van Gogh to the Netherlands. 

The Italian painter Leonardo Da Vinci learned painting in the workshop of Verrochio in Florence, 
and crafted there its first painting between 1472 and 1474. 
Da Vinci was based in Italy until 1516, when Francois I (King of France) invited him 
to join the Royal court as: "The King's First Painter, Engineer and Architect". 
Da Vinci died in France four years after his arrival.
Da Vinci's works was highly noted by French statesmen. Louis XII and (later) Napoleon though 
to bring ``The Last Supper" to France. ``Madonna of the Yarnwinder" is a painting done 
by Da Vinci to respond the demand of the secretary of state of Louis XII of France. 
Leonardo brought a version of the "Virgin of the Rocks" to France. One of the most important painting 
of Da Vinci is "Mona Lisa", currently displayed at Louvre Museum in Paris, was finalized 
in the Royal court of Francois I. All these elements about the relations between Da Vinci, 
France and Italy, explain the fact that Italy is strongly influenced by a link variation from 
Da Vinci to France, as shown in the bottom panel of Figure~\ref{fig:gohgvinci}.

\begin{figure}[!h]
\begin{center}
\includegraphics[scale=0.45]{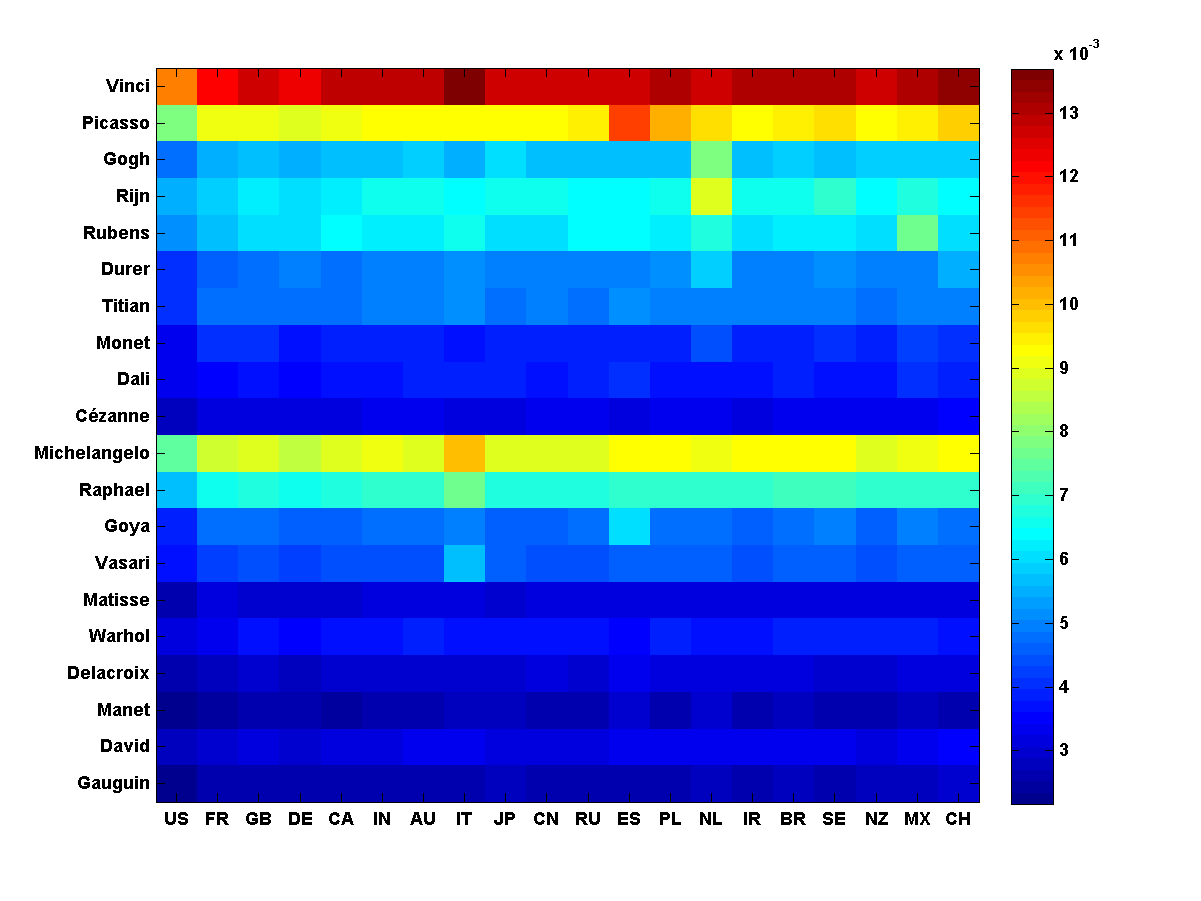}
\end{center}
\caption{{\bf Diagonal sensitivity of the top 20 countries to bidirectional link variations 
between painter/country pairs (i.~e.~painter to country and  country to painter)}
Color bar shows the sensitivity values. Data is averaged over 7 Wikipedia editions
and are shown for top 20 entries of 
Table~\ref{tab:painters40} and Table~\ref{tab:countries40}.
}
\label{fig:diagsensi}
%fig10paper
\end{figure}

\begin{figure*}[!h]
\begin{center}
\includegraphics[scale=0.25]{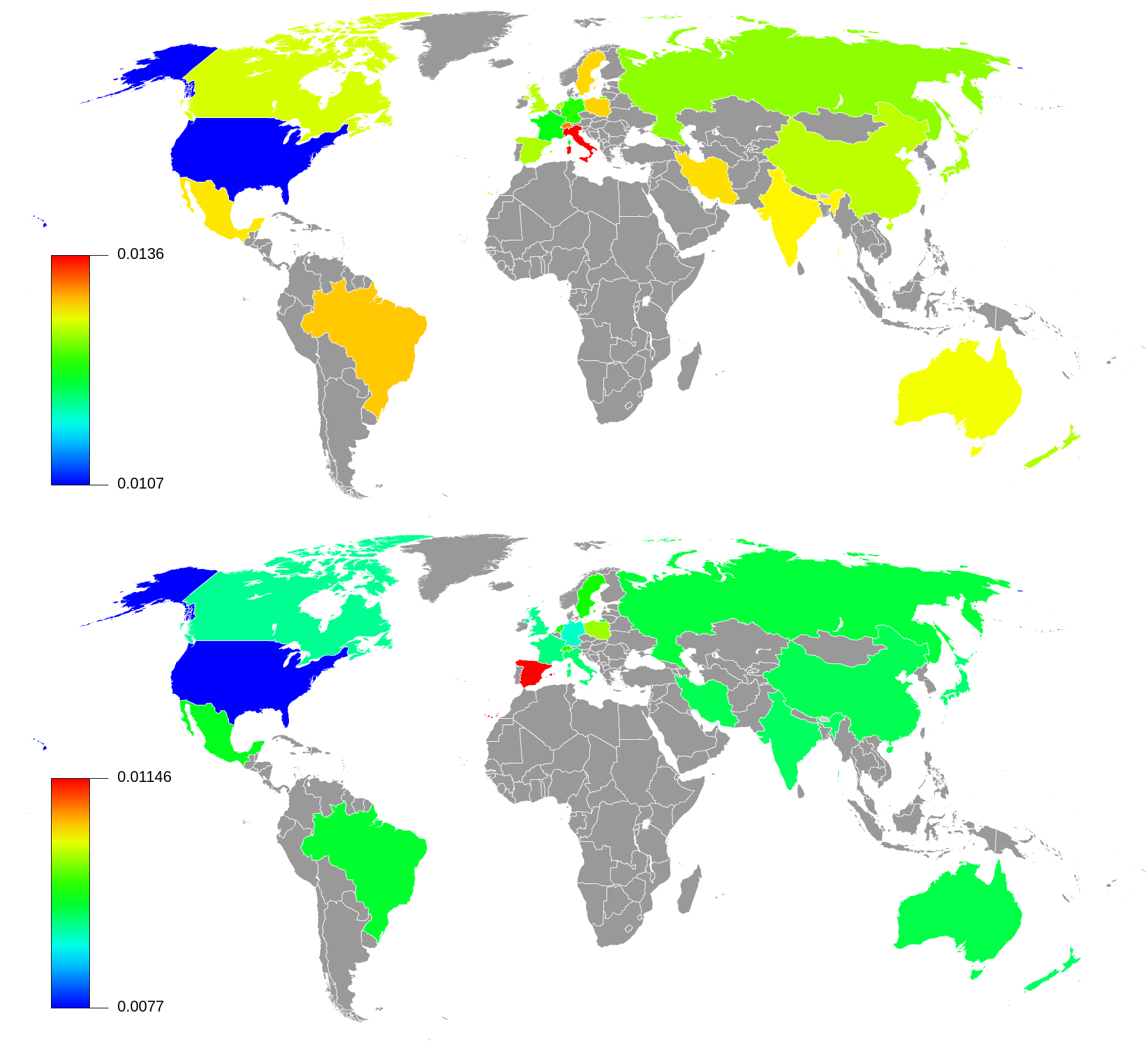}
\end{center}
\caption{{\bf Map representation of the diagonal sensitivity of countries to influence of 
Da Vinci  (top panel) 
and Picasso  (bottom panel).} 
}
\label{fig:vinci_picasso_world}
%fig11paper
\end{figure*}

{\it c) Diagonal sensitivity of countries:}
Finally in Figure~\ref{fig:diagsensi} we present the diagonal sensitivity of countries 
to their links with painters. This measure is computed by calculating the 2-way sensitivity 
of Eq.~\eqref{ep_2way} for each painter/country couple. It is the sum of 
the logarithmic PageRank sensitivity for the painter to country link and the one for the country to 
painter link. In Eq.~\eqref{ep_2way}, $c$ is the index of a country and $p$ of a painter.
\begin{equation} 
D_{(p \leftrightarrow c)}(c) = D_{(p \rightarrow c)}(c) + D_{(c \rightarrow p)}(c)
\label{ep_2way}
\end{equation}
Also, in Figure~\ref{fig:vinci_picasso_world}, we represent the same diagonal sensitivity 
of countries to Da Vinci and Picasso influence using a world map. In other words, we color the countries 
with the intensities found on the Da Vinci line (top panel) and the Picasso line (bottom panel) 
of the matrix of Figure~\ref{fig:diagsensi}. We have previously discussed the relationship between 
Picasso and Spain and the relation between Da Vinci and Italy
which are most sensitive countries in  Figure~\ref{fig:vinci_picasso_world} respectively. 
In this figure, we picture 
the secondly affected countries for each painter using the 2-way sensitivity metric for 
a the same bidirectional link variations. It is seen in  Figure~\ref{fig:vinci_picasso_world} (bottom)
that Poland is greatly impacted by Picasso. The reason is that 
the Mermaid of Warsaw is a symbol of Warsaw represented on city's 
coat as well as in a many imagery and statues. Picasso's drawing of Warsaw Mermaid explains 
the weight of 2-way sensitivity between Picasso and Poland and clarify why Poland is highly linked 
to Picasso \cite{picasso_poland}. On the other hand, 
from Figure~\ref{fig:vinci_picasso_world} (top) we find that
the second most influenced country by Da Vinci 
is Switzerland, possibly due to the fact that a central masterpiece of Da Vinci 
is to be found in Switzerland: Isabella d'Este \cite{vinci_switzerland}.

To finally conclude this analysis, we can underline that 
according to Figure~\ref{fig:diagsensi} Da Vinci, Picasso and Michelangelo 
are the most influential painters for selected world countries.

\section{Discussion}
\label{sec:conclu}
This paper shows that our sensitivity analysis captures the importance of relationships on network structure. 
This analysis relies on the reduced Google matrix and leverages its capability of concentrating 
all Wikipedia knowledge in a small stochastic matrix. We stress that the friendship networks and 
the sensitivity analysis of influence of painters on countries helped us extract valuable and realistic knowledge 
from a pure mathematical analysis without any direct appeal to arts, political, economical and social sciences. 
In a certain sense the reduced Google matrix approach provides an  artificial intelligence
analysis (the authors have no specific education in arts)
of interactions and influence of top painters on world countries using Wikipedia networks.

%%%%%%%%%%%%%%%%%%%%%%%%%%%%%%%%%%%%%%%%%%%

\section*{Acknowledgment}

This work was supported by APR 2015 call 
of University of Toulouse and by R\'egion Occitanie 
(project GOMOBILE), MASTODONS-2017 CNRS project APLIGOOGLE,
(see http://www.quantware.ups-tlse.fr/APLIGOOGLE/),
EU CHIST-ERA MACACO project 
ANR-13-CHR2-0002-06 (see http://macaco.inria.fr)
and in part by the Pogramme Investissements
d'Avenir ANR-11-IDEX-0002-02, reference ANR-10-LABX-0037-NEXT 
(project THETRACOM); 
it was granted access to the HPC resources of 
CALMIP (Toulouse) under the allocation 2017-P0110.
We would like to thank Dr. Hanah Tout from Lebanese University 
for her valuable comments and data validation.



%\bibliographystyle{IEEEtran}

\begin{thebibliography}{100L}

\bibitem{oxford} Oxford Dictionaries.
    https://en.oxforddictionaries.com/definition/art. Accessed Feb 2018
\bibitem{melivin} Melvin L.~Alexenberg. 
	{\it The Future of Art in a Digital Age: From Hellenistic to Hebraic Consciousness}, 
	Intellect Ltd (2008)
\bibitem{fandel} J.~Fandel. 
	{\it Pablo Picasso}, 
	Mankato, MN : Creative Education (2016)
\bibitem{richardson} J.~Richardson. 
	{\it A Life of Picasso. Vol. 1}, 
	 Random House (1991)
\bibitem{chipp}  Herschel B.~Chipp. 
	{\it Picasso's Guernica: History, Tranformations, Meanings}, 
	University of California Press (1988)
\bibitem{richardson2}  J.~Richardson. 
	{\it A Life of Picasso, Volume II: 1907-1917 - The Painter of Modern Life}, 
	Random House ebooks.
\bibitem{wolf}  N.~Wolf. 
	{\it Expressionism}, 
	Taschen (2004).
\bibitem{benson}  Timothy O.~Benson,  L.~Easton, C.~Grammont and  F.~Josenhans.
	{\it Expressionism in Germany and France: From Van Gogh to Kandinsky}, 
	LACMA (2014).
\bibitem{wikiorg} http://www.wikipedia.org (Accessed Feb 2018)
\bibitem{britanica} Encyclopaedia Brittanica 
          http://www.britannica.com/ (Accessed Feb 2018).
\bibitem{giles} J.Giles, {\it Internet encyclopaedias go head to head},
               Nature {\bf 438}, 900 (2005)
\bibitem{reagle} J.M.~Reagle Jr. {\it  Good Faith Collaboration: 
              The Culture of Wikipedia},
                MIT Press, Cambridge MA (2010)
% ISBN 978-0-262-01447-2
\bibitem{finn} F.A.~Nielsen, {\it Wikipedia research and tools: 
         review and comments}, (2012),
         available at SSRN: 
         {\footnotesize \verb|dx.doi.org/10.2139/ssrn.2129874|}
\bibitem{brin} S.~Brin and L.~Page,
         {\it The anatomy of a large-scale hypertextual Web search engine},
         Computer Networks and ISDN Systems {\bf 30}, 107 (1998)
\bibitem{meyer} A.M.~Langville and C.D.~Meyer,
         {\it Google's PageRank and beyond: 
         the science of  search engine rankings},
         Princeton University Press, Princeton (2006)
\bibitem{rmp2015} L.~Ermann, K.M.~Frahm and D.L.~Shepelyansky,
        {\it Google matrix analysis of directed networks},
          Rev. Mod. Phys. {\bf 87}, 1261 (2015)
\bibitem{wikizzs} A.O.~Zhirov, O.V.~Zhirov and D.L.~Shepelyansky,
        {\it Two-dimensional ranking of Wikipedia articles},
        Eur. Phys. J. B {\bf 77}, 523 (2010)
\bibitem{wikievol} Y.-H.~Eom, K.M.~Frahm, A.~Benczur and D.L.~Shepelyansky, 
        {\it Time evolution of Wikipedia network ranking}, 
         Eur. Phys. J. B {\bf 86}, 492 (2013)
\bibitem{eomwiki9} Y.-H.~Eom and D.L.~Shepelyansky,
        {\it Highlighting entanglement of cultures 
         via ranking of multilingual Wikipedia articles},
         PLoS ONE {\bf 8(10)},  e74554 (2013) 
\bibitem{eomwiki24} Y.-H.~Eom, P.Aragon, D.Laniado, 
          A.Kaltenbrunner, S.Vigna and D.L.~Shepelyansky,
         {\it Interactions of cultures and top people of 
         Wikipedia from ranking of 24 language editions},
         PLoS ONE {\bf 10(3)}, e0114825 (2015)
\bibitem{lageswiki} J.Lages, A.Patt and D.L.Shepelyansky, 
        {\it Wikipedia ranking of world universities}, 
        Eur. Phys. J. B {\bf 89}, 69 (2016)
\bibitem{rokach} G.~Katz and L.~Rokach,
         {\it Wikiometrics: a Wikipedia based ranking system},
         World Wide Web {\bf 20(6)}, 1153 (2017)
%        DOI:10.1007/s11280-016-0427-8 Springer N.Y.
\bibitem{hart} M.H.~Hart, {\it The 100:  ranking of the most 
               influential persons in history}, Citadel Press, N.Y. (1992)
\bibitem{shanghai} Academic Ranking of World Universities,
           http://www.shanghairanking.com/ 
           (Accessed Feb 2018)
\bibitem{vigna} R.~Meusel, S.~Vigna, O.~Lehmberg  and  C.~Bizer,
                {\it  The graph structure in the web analyzed on different aggregation levels},
                 J. Web Sci. {\bf 1}, 33 (2015)
%%             doi: 10.1145/2567948.2576928doi
%\bibitem{linux}  A.D.~Chepelianskii,
%         {\it Towards physical laws for software architecture},
%         arXiv:1003.5455 [cs.SE] (2010)
\bibitem{greduced} K.M.~Frahm and D.L.~Shepelyansky,
         {\it Reduced Google matrix},
         arXiv:1602.02394[physics.soc] (2016)
\bibitem{politwiki} K.M.~Frahm, K.~Jaffr\`es-Runser and D.L.~Shepelyansky,
         {\it Wikipedia mining of hidden links between political leaders},
          Eur. Phys. J. B {\bf 89}, 269 (2016) 
\bibitem{geop} K.M.~Frahm, S.~El~Zant, K.~Jaffr\`es-Runser, D.L.~Shepelyansky,
 	{\it Multi-cultural Wikipedia mining of geopolitics interactions 
        leveraging reduced Google matrix analysis},
	Phys. Lett. A \textbf{381}, 2677 (2017)
\bibitem{menacomm} 
 S.~El~Zant, K.M.~Frahm, K.~Jaffr\`es-Runser, D.L.~Shepelyansky,
 	{\it Geopolitical interactions from reduced Google matrix analysis of Wikipedia},
Proceedings of IEEE MENACOMM 2018, Journieh, Lebannon, April 2018.
\bibitem{allnames}
	https://en.wikipedia.org/wiki/List\_of\_painters\_by\_name (Accessed Oct 2017)
\bibitem{ourwebpage} http://www.quantware.ups-tlse.fr/QWLIB/wikipainternets/ 
          (Accessed Feb 2018).
\bibitem{picassobraque}
    {\it Georges Braque and Pablo Picasso}.
    https://www.masterworksfineart.com/blog/georges-braque-and-pablo-picasso/ 
    (Accessed Feb 2018).
\bibitem{pissaromonet} 
    {\it Monet and Pissarro in London}.
    http://www.visual-arts-cork.com/history-of-art/claude-monet-camille-pissarro-in-london.htm
    (Accessed Feb 2018).
\bibitem{romatisse1}
    {\it Rouault / Matisse, correspondances au mus\'{e}e d'Art moderne de la Ville de Paris}.
    http://lucileee.blog.lemonde.fr/2007/01/08/rouault-matisse-correspondances-au-musee-dart-moderne-de-la-ville-de-paris/ (Accessed Feb 2018).
\bibitem{romatisse2}
    {\it Matisse-Rouault : correspondance 1906-1953: une vive sympathie d'art.}
    https://critiquedart.revues.org/13275 (Accessed Feb 2018).
\bibitem{schneider}  P.~Schneider
	{\it Matisse}, 
	Flammarion (2002).
\bibitem{wildenstein}  D.~Wildenstein.
	{\it Monet or The Triumph of Impressionism}, 
	Taschen (2014).
\bibitem{montebello}  P.~Montebello, J.N.~Wood, D.~Wildenstein and C.S.~Moffett
	{\it Monet's Years at Giverny: Beyond Impressionism}, 
	Harry N.Abrams, INC, Publisher New York (1995).
\bibitem{mich}
    {\it Michelangelo and Leonardo Da Vinci}.
    https://www.michelangelo.org/michelangelo-and-da-vinci.jsp (Accessed Feb 2018).
%\newpage
\bibitem{berger}  J.~Berger.
	{\it The Success and Failure of Picasso}, 
	Vintage (1993).
\bibitem{dali} Independent.
    {\it Picasso, Mir\'{o}, Dal\'{i}: The Birth of Modernity}.
    http://www.independent.co.uk/arts-entertainment/art/reviews/picasso-mir-dal-the-birth-of-modernity-palazzo-strozzi-florence-picasso-in-paris-van-gogh-museum-2254050.html 
    (Accessed Feb 2018).
\bibitem{gephi} M. Bastian, S. Heymann, M. Jacomy.
	{\it Gephi: An Open Source Software for Exploring and Manipulating Networks}.
	Proc. of International AAAI Conference on Weblogs and Social Media (2009).
%\bibitem{taylor}   M.R.~Taylor, A.~Lins, M.S.~Meighan, B.A.~Price, K.~Sutherland, S.~Homolka and E.~Torok
%	{\it Marcel Duchamp: \'{e}tant donn\'{e}s}, 
%	Yale University Press (2009).
%\bibitem{venezia}   M.~Venezia
%	{\it Henri de Toulouse-Lautrec (Getting to Know the World's Greatest Artists)}, 
%	Childrens Press, Chicago (1995).
%\bibitem{chtchoukine} 
%	\url{http://www.collectionchtchoukine.com}.
\bibitem{ferrier}   J.L.~Ferrier
	{\it The Fauves: The Reign of Colour}, 
	Terrail (1995).
\bibitem{brodskaia}   N.~Brodskaia
	{\it The Fauves : The Masters Who Shook the World of Art (Schools \& Movements Series)}, 
	Parkstone Press (1996).
\bibitem{whitfield}  S.~Whitfield
	{\it Fauvism (World of Art)}, 
	Thames \& Hudson (1996).
\bibitem{antliff}  M.~Antliff
	{\it Cubism and Culture}, 
	Thames \& Hudson (2001).
\bibitem{wikiiso2} https://en.wikipedia.org/wiki/ISO\_3166-1\_alpha-2 (Accessed Feb 2018).
\bibitem{spgroups} S.~El Zant, K.M.~Frahm, K.Jaffres-Runser and D.L.Shepelyansky, 
         {\it Analysis of world terror networks from the reduced Google matrix of Wikipedia},
         Eur. Phys. J. B {\bf 91}, 7 (2018).

\bibitem{picasso_poland}
    {\it The Matador \& the Mermaid: A Story of Picasso \& World Peace}.
    https://culture.pl/en/article/the-matador-the-mermaid-a-story-of-picasso-world-peace (Accessed May 2018).
\bibitem{vinci_switzerland}
    {\it Leonardo Da Vinci painting lost for centuries found in Swiss bank vault}.
    The telegraph,
    https://goo.gl/tLjrxC (Accessed May 2018).
     
\end{thebibliography}

\begin{IEEEbiography}[{\includegraphics[width=1in,height=1.25in,clip,keepaspectratio]{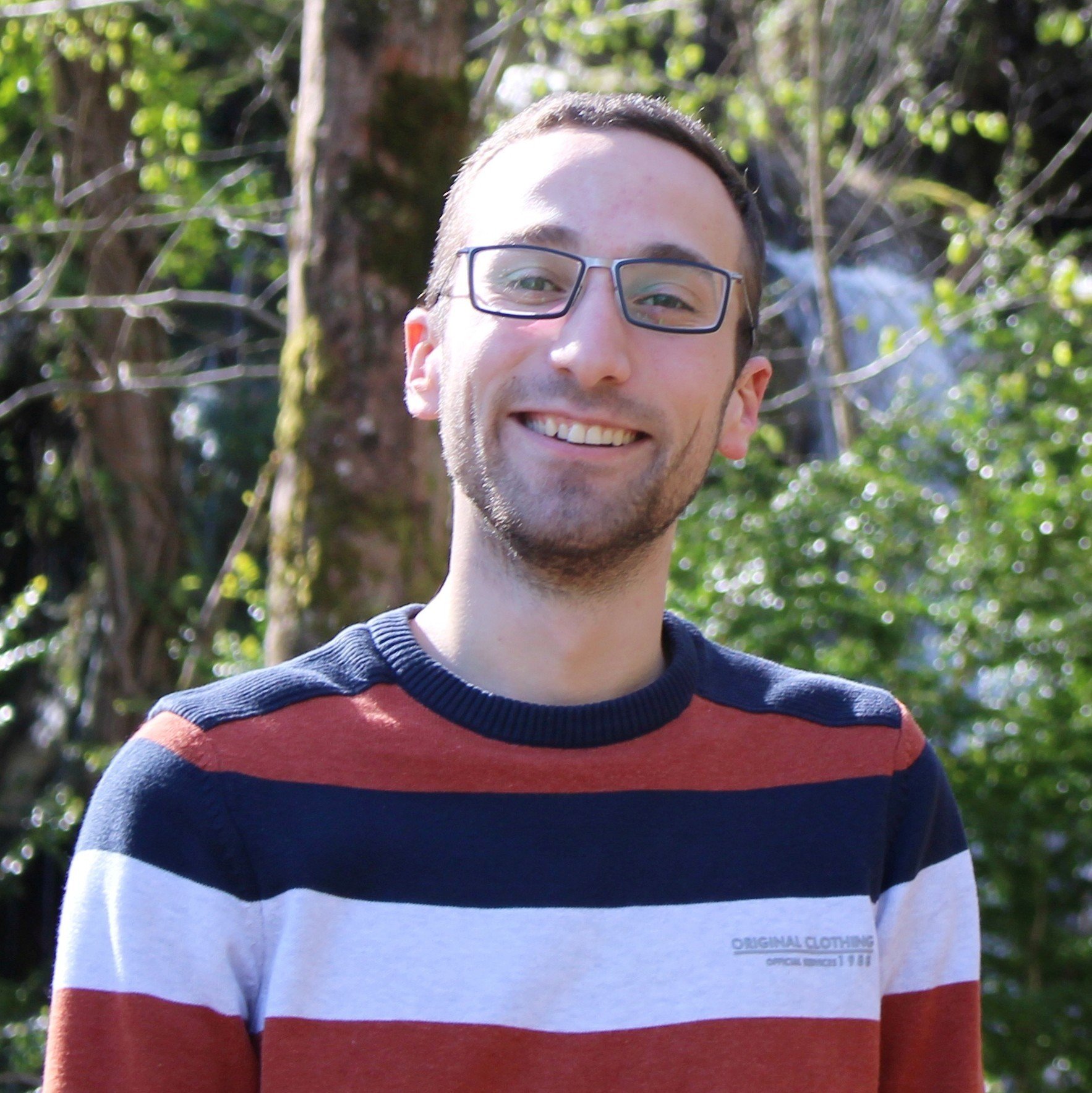}}]{Samer El Zant}
received an engineering degree (B.E.) in Communications and electronics from Beirut University in 2014 
and a Master of engineering in Networks and systems from the University of Versailles, in 2015. 
He is currently an instructor at \'{E}cole nationale sup\'{e}rieure d'\'{e}lectrotechnique, d'\'{e}lectronique, 
d'informatique, d'hydraulique et des t\'{e}l\'{e}communications (ENSEEIHT, Toulouse), 
and a PhD student at Institut national polyt\'{e}chnique de Toulouse (INPT, Toulouse). 
His researches focus on reduced Google matrix analysis on Big Data as Wikipedia, Twitter and genetic networks.
\end{IEEEbiography}

\vfill

\begin{IEEEbiography}[{\includegraphics[width=1in,height=1.25in,clip,keepaspectratio]{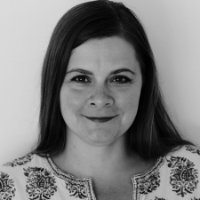}}]{Katia Jaffr\`es-Runser}
received both a Dipl. Ing. (M.Sc.) in Telecommunications and a DEA (M.Sc) in Medical Imaging in 2002 and a Ph.D. in Computer Science in 2005 from the National Institute of Applied Sciences (INSA), Lyon, France. 
From 2002 to 2005 she was with Inria, participating in the ARES project while working towards her Ph.D. thesis. 
In 2006, she joined the Stevens Institute of Technology, Hoboken, NJ, USA, as a post-doctoral researcher.
She is the recipient of a three-year Marie-Curie OIF fellowship from the European Union to pursue her work from 2007 to 2010. 
She currently holds a Ma\^itre de conf\'erences (Associate Professor) at University of Toulouse, Toulouse INP-ENSEEIHT. She's a member of the IRIT laboratory. 
She's interested in the performance evaluation of networks in general. 
\end{IEEEbiography}

\vfill

\begin{IEEEbiography}[{\includegraphics[width=1in,height=1.25in,clip,keepaspectratio]{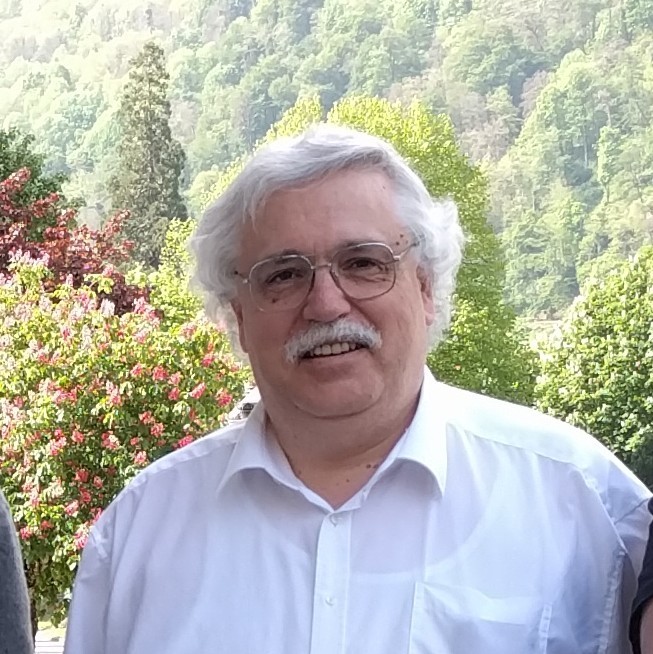}}]{Klaus M. Frahm}
received a Diplom degree in Physics in 1989 and his PhD in 
1993 at the University of Cologne. After being a post-doctoral researcher at the CEA in Saclay 
and the University of Leiden he came to the University Paul Sabatier in 
Toulouse where he obtained in 1998 his HDR degree and is 
full professor since 2000. Currently he is member of the 
Laboratoire de Physique Th\'eorique du CNRS in Toulouse (LPT, UMR 5152). 
His research focuses on random matrix theory with applications to 
chaotic scattering and quantum transport, localization and interactions, 
and  Perron-Frobenius operators with applications to 
Google matrices of directed networks.
\end{IEEEbiography}

\vfill

\begin{IEEEbiography}[{\includegraphics[width=1in,height=1.25in,clip,keepaspectratio]{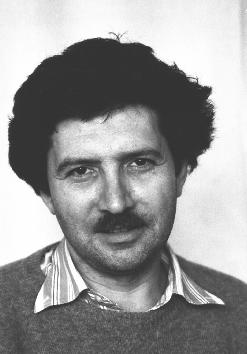}}]{Dima L. Shepelyansky}
received a Diplom degree in Physics in 1978 at Novosibirsk State University,
PhD in 1982 and Russian Doctor Degree in 1989
at Institute of Nuclear Physics of Russian Academy of Sciences, 
Novosibirsk, Russia.
He is a researcher at CNRS Toulouse, France from 1991 and Directeur de Recherche from 1994 
at  Laboratoire de Physique Th\'eorique du CNRS, Toulouse.
His research fields include classical and quantum chaos, 
nonlinear waves, dark matter dynamics,
atom ionization in strong fields, electronic transport, 
quantum computing, Markov chains and Google matrix of directed networks.
\end{IEEEbiography}

\vfill

%\EOD

\end{document}